\documentclass[onecolumn,usenatbib]{mn2e}

\usepackage{amsmath}
\usepackage[dvips]{graphicx}
\usepackage{verbatim}

\newcommand{\prob}{\textrm{Pr}}
\newcommand{\dif}{\mathrm{d}}

\newcommand{\bs}{\boldsymbol}
\newcommand{\trans}{\mathrm{T}}
\newcommand{\trace}{\textrm{Tr}}
\newcommand{\expect}{\mathrm{E}}

\begin{document}

\title
[Simulations of partially coherent imaging arrays]
{Simulations of 
partially coherent focal plane imaging arrays:
Fisher matrix approach to performance evaluation}

\author
[Saklatvala et al.]
{G.~ Saklatvala,\thanks{Email: g.saklatvala@mrao/cam.ac.uk} 
S.~Withington,\thanks{Email: stafford@mrao.cam.ac.uk} 
and M.~Hobson\thanks{Email: mph@mrao.cam.ac.uk} \\
Cavendish Laboratory, J.J.~Thomson Avenue, Cambridge CB3 0HE, UK}

\date{Submitted to MNRAS 7 March 2007}

\maketitle

\begin{abstract}
Focal plane arrays of bolometers are increasingly employed in astronomy 
at far--infrared to millimetre wavelengths. The focal plane
fields and the detectors are both partially coherent in these systems,
but no account has previously been taken of the effect of partial
coherence on array performance. In this paper, we use our recently
developed coupled--mode theory of detection together with Fisher information
matrix techniques from signal processing to characterize the behaviour
of partially coherent imaging arrays. We investigate the effects 
of the size and coherence length of both the source and the detectors, and the
packing density of the array, 
on the amount of information that can be extracted from observations with such
arrays.
\end{abstract}


\section{Introduction}

Large format focal plane imaging arrays are of increasing importance in many
areas of submillimetre wave astronomy. 
Arrays utilising semiconducting bolometers
have already been built for 
the Stratospheric Observatory For Infrared Astronomy 
(SOFIA) \citep{evans2002:1} and SHARC II \citep{dowell2003:1}, each of which 
utilises $12 \times 32$ arrays. Future arrays are planned
using transition edge sensors, including SCUBA 2 \citep{holland2006:1},
GISMO \citep{staguhn2006:1}
and arrays for the Green Bank Telescope (GBT) \citep{dicker2006:1},
Atacama Cosmology Telescope (ACT) \citep{fowler2004:1} and CLOVER
\citep{maffei2005:1}. The largest
of these planned arrays is of size $64 \times 64$ pixels.

The construction of imaging arrays such as these, all of which operate
at wavelengths between 0.05mm and 3mm, presents difficulties
in understanding their behaviour. Crucially, no account has previously
been taken
of the effects of coherence on the performance of an array. This is important
for two reasons. Firstly, at the wavelengths concerned telescopes
are few moded, resulting in a partially coherent field in the
focal plane, with correlations between different pixels. 
Secondly, the detectors used in the arrays are only a few 
wavelengths in size (in the case of ACT less than a wavelength), and
should be expected therefore to show partially coherent behaviour themselves.
If partially coherent imaging arrays are to be used successfully for
astronomical observations,
it is essential to understand their behaviour. 

In the past, understanding of partially coherent detectors  
was limited by the lack of practical modelling techniques. 
In a recent paper \citep*{saklatvala2007:1}
we developed a general theory of partially coherent detection; in this
paper we apply the technique to the simulation of imaging arrays.
Qualitatively, coherence has two effects: it increases the
noise associated with the statistical variation of the
incoming field, and leads to correlations in this noise. We employ a
Fisher matrix analysis to gain insight into how these effects alter array 
performance. In our simulations, we compare the ability of different
arrays to recover information about some simple sources. Specifically,
we focus on the size and coherence both of the detectors and the sources,
and the packing density of the array. We anticipate that such analyses will
prove useful in the design of the next generation of astronomical
focal plane bolometric imaging arrays.

\section{\label{theory}Theory of partially coherent detectors}
In earlier papers \citep*{withington2001:1, withington2002:1,saklatvala2007:1}
we derived the following expression for the average power
absorbed by a paraxial multimode power detector from a statistically
stationary source:
\begin{eqnarray}\label{powercont}
\langle P \rangle  &=& \int_0^\infty \dif \omega \sum_{i=1}^2 \sum_{j=1}^2
\int_\mathcal{S} \dif \bs{r}_1 \int_\mathcal{S}
\dif \bs{r}_2 
Z^*_{ij}(\bs{r}_1,\bs{r}_2,\omega) Y_{ij}(\bs{r}_1,\bs{r}_2,\omega) \;,
\end{eqnarray}
where $Y_{ij}(\bs{r}_1,\bs{r}_2,\omega)$ is the cross-spectral density
\begin{eqnarray}\label{csd}
Y_{ij}(\bs{r}_1,\bs{r}_2,\omega) &=& \int_{-\infty}^\infty e^{-i\omega u}
\langle E_i(\bs{r}_1,t) E^*_j(\bs{r}_2,t+u) \rangle  \dif u \;,
\end{eqnarray}
and
$E_i(\bs{r}_1,t)$ is the $i$th component of the analytic signal
associated with the incoming electric field at 
position $\bs{r}_1$ and time $t$.
$\mathcal{S}$ is a surface normal to the wavevector characterizing
the input of the detector, and the
sums are over the two polarizations normal to the wavevector.
$Z_{ij}(\bs{r}_1,\bs{r}_2,\omega)$ is a quantity characterizing the
detector which we call the `detector coherence tensor'.
We have derived (\ref{powercont}) in several ways, using thermodynamic 
arguments \citep{withington2001:1}, reciprocity \citep{withington2002:1},
and general arguments about the properties of detectors 
\citep{saklatvala2007:1}. The result is effectively a weighted linear
sum of all pairs of space--time correlations in the incoming field, 
and may be interpreted also in terms of the coupling between the natural modes
of the incoming field and a set of modes that characterizes the detector.
The power of our theory is that, in addition to calculating the average
detector outputs, it also allows us to relate the statistics of the
incoming fields to the statistics of the detector outputs.
For a thermal source, for an integration time
much longer than either the coherence time of the source or the response time
of the detector,
the covariance of the power recorded by two
detectors is
\begin{eqnarray}\label{covcont}
\textrm{Cov}[P^a,P^b] &=& \frac{1}{\tau}
\int_0^\infty \dif \omega \sum_{i=1}^2 \sum_{j=1}^2
\sum_{k=1}^2 \sum_{l=1}^2 
\int_\mathcal{S} \dif \bs{r}_1 \int_\mathcal{S}
\dif \bs{r}_2 \int_\mathcal{S} \dif \bs{r}_3 \int_\mathcal{S}
\dif \bs{r}_4 
Z^{a*}_{ji}(\bs{r}_2,\bs{r}_1,\omega) Y_{jk}(\bs{r}_2,\bs{r}_3,\omega)
Z^{b*}_{lk}(\bs{r}_4,\bs{r}_3,\omega) Y_{li}(\bs{r}_4,\bs{r}_1,\omega) 
\nonumber \\ &&
+ \frac{\delta_{ab}}{\tau} \int_0^\infty \dif \omega \sum_{i=1}^2 \sum_{j=1}^2
\int_\mathcal{S} \dif \bs{r}_1 \int_\mathcal{S}
\dif \bs{r}_2 \hbar \omega 
Z^{a*}_{ij}(\bs{r}_1,\bs{r}_2,\omega) Y_{ij}(\bs{r}_1,\bs{r}_2,\omega)
\;,
\end{eqnarray}
where $\tau$ is the integration time.
(\ref{covcont}) can be derived using either a semi--classical 
\citep{saklatvala2007:1} or quantum optical \citep*{zmuidzinas2003:1, 
withington2005:1} approach. (\ref{covcont}) gives both the fluctuations of 
the detector outputs and the correlations between
the outputs, and thus incorporates the Hanbury Brown--Twiss effect 
\citep{hanburybrown1956:1}.

In most applications, a frequency filter is placed in front of a detector
to restrict the bandwidth (this is necessary to avoid blurring of the
image, in the case of an array, or ambiguity in the baselines, 
in the case of an interferometer). As an idealization, we consider
a filter that multiplies the cross--spectral density by a top
hat function with central frequency $\omega_0$ and bandwidth $\Delta\omega$.
To avoid blurring, the bandwidth must be sufficiently small that
the cross-spectral density and the detector coherence tensor
are approximately constant across the bandwidth of the detector, in which case
\begin{eqnarray}\label{narrowpowercont}
\langle P \rangle &\simeq&  
\Delta \omega
\sum_{i=1}^2 \sum_{j=1}^2
\int_\mathcal{S} \dif \bs{r}_1 \int_\mathcal{S}
\dif \bs{r}_2 
Z^*_{ij}(\bs{r}_1,\bs{r}_2,\omega_0) Y_{ij}(\bs{r}_1,\bs{r}_2,\omega_0) \;,
\end{eqnarray}
and
\begin{eqnarray}\label{narrowcovcont}
\textrm{Cov}[P^a,P^b] &\simeq& 
\frac{\Delta\omega}{\tau} \sum_{i=1}^2 \sum_{j=1}^2
\sum_{k=1}^2 \sum_{l=1}^2 
\int_\mathcal{S} \dif \bs{r}_1 \int_\mathcal{S}
\dif \bs{r}_2 \int_\mathcal{S} \dif \bs{r}_3 \int_\mathcal{S}
\dif \bs{r}_4 
Z^{a*}_{ji}(\bs{r}_2,\bs{r}_1,\omega_0) Y_{jk}(\bs{r}_2,\bs{r}_3,\omega_0)
Z^{b*}_{lk}(\bs{r}_4,\bs{r}_3,\omega_0) Y_{li}(\bs{r}_4,\bs{r}_1,\omega_0) 
\nonumber \\ &&
+ \frac{\hbar\omega_0\Delta\omega\delta_{ab}}{\tau} \sum_{i=1}^2 \sum_{j=1}^2
\int_\mathcal{S} \dif \bs{r}_1 \int_\mathcal{S}
\dif \bs{r}_2 
Z^{a*}_{ij}(\bs{r}_1,\bs{r}_2,\omega_0) Y_{ij}(\bs{r}_1,\bs{r}_2,\omega_0) \;.
\end{eqnarray}
For numerical work we must discretize the integrals in (\ref{narrowpowercont})
and (\ref{narrowcovcont}). The idealized detectors we simulate in this paper
respond equally to all polarizations, so defining
\begin{eqnarray} 
\mathrm{Z}_{\alpha\beta} &=& (\Delta x)^2 
Z_{ij}(\bs{r}_\alpha,\bs{r}_\beta,\omega_0) \label{zdisc}
\\
\mathrm{Y}_{\alpha\beta} &=& \sum_{ij} 
Y_{ij}(\bs{r}_\alpha,\bs{r}_\beta,\omega_0) \label{ydisc}
\end{eqnarray}
for some regular grid of sample points $\{\bs{r}_\alpha\}$
with spacing $\Delta x$, we obtain
\begin{eqnarray} \label{power}
\langle P \rangle &\sim& \Delta\omega \trace\mathbf{Z} \mathbf{Y} \;,
\end{eqnarray}
and
\begin{eqnarray} \label{narrowbandcov}
\textrm{Cov}[P^a,P^b] &\sim& \frac{\Delta\omega}{\tau} \left(
\trace\mathbf{Z}^a\mathbf{YZ}^b\mathbf{Y}
+ \delta^{ab}\hbar \omega_0 \trace\mathbf{Z}^a\mathbf{Y}  \right) \;.
\end{eqnarray}
Note that we have used italic letters to denote the continuous tensors
and Roman letters to denote the discretized forms.

In the single mode case, the
signal--to--noise ratio $P/\sqrt{\textrm{Cov}[P,P]}$ obtained
from (\ref{power}) and (\ref{narrowbandcov})
reduces to the Dicke radiometer equation \citep{dicke1946:1}
in the high occupancy limit,
and Poisson noise in the low occupancy limit.
In this paper we are interested in the long wavelength, high occupancy
limit, so we neglect the second term of (\ref{narrowbandcov}).

\section{Parameter estimation from imaging arrays}
A crucial factor in measuring the performance of an imaging array is
the amount of information that
can be extracted from it. 
To this end, we employ the technique of Fisher matrix analysis, which is widely
used in many areas of astronomy (\citealt*{tegmark1997:1,hamilton1997:1}; 
\citealt{yamamoto2001:1,martins2002:1,rocha2004:1}),
particularly the constraint of cosmological parameters
from CMB measurements, and has also been applied to the design of radio
interferometry arrays \citep{stoica1989:1,abramovich1996:1,adorf1996:1},
but has not hitherto to our knowledge been employed in
the context of focal plane imaging arrays. 
The Fisher matrix analysis technique provides a way of estimating the 
uncertainties and correlations in the parameters
of a model fitted to some data, 
averaged over all realisations of the data, and without the
need to enumerate specific realisations. In this paper we perform
simulations to estimate the constraints on the parameters of a source
when observed with imaging arrays of different size, coherence and packing
density.
 
Given a data column vector $\bs{D}$ and a model $\mathcal{H}$ 
with parameter column vector $\bs{\theta}$,
the Fisher information matrix is 
equal to minus the expected 
curvature of the logarithm of the
likelihood:
\begin{eqnarray}\label{fisherdef}
F_{ij}[\bs{\theta},\mathcal{H}] &=& - \expect\left[\frac{\partial^2}
{\partial\theta_i\partial\theta_j} \ln\prob(\bs{D}|\bs{\theta},\mathcal{H})
\right] \;.
\end{eqnarray}
For a uniform prior, the posterior is proportional to the likelihood,
so the Fisher matrix is also equal to minus the expected 
curvature of the logarithm of the
posterior.
Applying the Gaussian approximation
to the posterior, we see that the covariance matrix of the recovered parameters
can be estimated by inverting the Fisher matrix for the parameter values
at the maximum of the posterior. Indeed it can be shown 
(\citealt{marzetta1993:1}, and
see Appendices \ref{APPA} and \ref{APPB}) that the inverse of the 
Fisher matrix provides a lower bound for the covariance matrix
of any unbiased estimators
of the parameters (the Cram\'er--Rao bound).

The Fisher matrix at the maximum of the posterior
can itself be estimated from the mean and covariance
of the data, and the derivatives of these quantities with respect to the
parameters. Applying the Gaussian approximation to the likelihood
function, we can show (see Appendix \ref{APPC}) that
\begin{eqnarray}\label{fisherapprox}
F_{ij}[\bs{\theta}_0,\mathcal{H}] &\simeq&
\frac{\partial\bs{\mu}^\trans[\bs{\theta}_0,\mathcal{H}]}{\partial\theta_i}
\bs{\Sigma}^{-1}[\bs{\theta_0}]
\frac{\partial\bs{\mu}[\bs{\theta_0}]}{\partial\theta_j}
+
\frac{1}{2} \trace\left(
\bs{\Sigma}^{-1}[\bs{\theta}_0,\mathcal{H}] \frac{\partial\bs{\Sigma}[\bs{\theta}_0,\mathcal{H}]}
{\partial\theta_i}
\bs{\Sigma}^{-1}[\bs{\theta}_0,\mathcal{H}] \frac{\partial\bs{\Sigma}[\bs{\theta}_0,\mathcal{H}]}
{\partial\theta_j}
\right) \;,
\end{eqnarray}
where
\begin{eqnarray}
\bs{\mu}[\bs{\theta},\mathcal{H}] &\equiv& \expect[\bs{D}] \equiv
\int \prob(\bs{D}|\bs{\theta},\mathcal{H}) \bs{D} \dif \bs{D}
\end{eqnarray}
and
\begin{eqnarray}
\bs{\Sigma}[\bs{\theta},\mathcal{H}] &\equiv& 
\expect[(\bs{D}-\bs{\mu}[\bs{\theta},\mathcal{H}])
(\bs{D}-\bs{\mu}[\bs{\theta},\mathcal{H}])^\trans] \equiv
\int \prob(\bs{D}|\bs{\theta},\mathcal{H})
(\bs{D}-\bs{\mu}[\bs{\theta},\mathcal{H}])
(\bs{D}-\bs{\mu}[\bs{\theta},\mathcal{H}])^\trans \dif \bs{D}
\end{eqnarray}
are evaluated at $\bs{\theta} = \bs{\theta}_0$,
satisfying
\begin{eqnarray}
\left.\expect\left[\frac{\partial}{\partial\theta_i}
\prob(\bs{D}|\bs{\theta},\mathcal{H})\right]\right|_{\bs{\theta}=\bs{\theta}_0}
&=& 0 \;.
\end{eqnarray}

To model an imaging array, we use the results (\ref{power}) and 
(\ref{narrowbandcov}) and set
\begin{eqnarray}
D_i &=& \langle P^i \rangle \;, \label{datavector}\\
\Sigma_{ij} &=& \textrm{Cov}[P^i,P^j] \label{covmatrix}\;,
\end{eqnarray}
where $P^i$ is the output of the $i$th detector.
With this substitution it can be seen that the first term of 
(\ref{fisherapprox}) is proportional to $\tau\Delta\omega$, and therefore
proportional to the ratio of integration time to coherence time;
whereas the second term is independent of this ratio.
For (\ref{narrowbandcov}) to hold the integration time must be much greater
than the coherence time, and hence we can ignore the second term
of (\ref{fisherapprox}). 
We see that the covariance of the
recovered parameters scales with the reciprocal of the number of 
coherence times per integration time.
As the ratio of integration time 
to coherence time tends to infinity, the uncertainty in the parameters
therefore tends to zero, but in reality, the integration time is limited by
practical factors. 

Note that although the final result is the same as in previous
astronomical applications, the physics is significantly different:
the fluctuations and correlations depend on the parameters of the model,
whereas in previous work the noise terms have been independent of the 
parameters. It is clearly possible to add parameter independent noise
terms to (\ref{narrowbandcov}), but since the purpose of this paper
is to demonstrate the effect of coherence on imaging arrays,
we ignore these additional terms.

\section{Parametrization of sources and detectors}
In this paper we perform simulations to investigate the ability of
different imaging arrays to extract information from a source. We
therefore need suitable parametrizations for both the sources and the 
detectors.
\subsection{Parametrization of sources}
The performance of an imaging array will clearly depend on the nature
of the sources being observed; there is no universal measure
of absolute performance. We can however gain considerable insight by
investigating the ability of imaging arrays with different coherence
lengths and geometries to determine the parameters of a Gaussian--Schell
source \citep{collett1978:1}:
\begin{eqnarray}\label{gauss-schell}
Y_{ij}(\bs{r}_1,\bs{r}_2) &=& \frac{I}{\sqrt{2\pi\sigma_\textrm{s}^2}}
\exp\left(-\frac{|\bs{r}_1-\bs{r}_\textrm{c}|^2 + |\bs{r}_2-\bs{r}_\textrm{c}|^2}{4\sigma_\textrm{s}^2}
-\frac{|\bs{r}_1-\bs{r}_2|^2}{2\sigma_\textrm{g}^2}\right) \;,
\end{eqnarray}
where $\bs{r}_\textrm{c}\equiv(x_\textrm{c},y_\textrm{c})$ is the position of the source, $I$ is the total
power, $\sigma_\textrm{s}$ is the geometrical width and $\sigma_\textrm{g}$ is the coherence
length. There are several advantages to using such a source in our
simulations. Firstly, it is often a good approximation to the
coherence function in the focal plane of a telescope when either
a point source or incoherent Gaussian source is present on the sky. 
Secondly, there is a clear
modal interpretation: the natural modes of a Gaussian--Schell source
are the Gauss-Hermite modes.
Thirdly, there are 
sufficiently few parameters that the Fisher matrices generated can
be easily interpreted. There is no explicit reference to wavelength in
the Gaussian--Schell parametrization, but in practise the wavelength dependence
is implicit in the coherence length and geometrical size of the source.

In the simulations we use the parameter vector
\begin{eqnarray}\label{paramvector}
\bs{\theta} = (x_\textrm{c},y_\textrm{c},I,\sigma_\textrm{s})^\mathrm{T} \;.
\end{eqnarray}  
The coherence length $\sigma_g$ is not included as a parameter to be 
recovered: our simulations verified
that it is highly degenerate with the intensity, and thus leads
to a poorly conditioned Fisher matrix. As the coherence length of the 
incoming field in the focal plane 
depends primarily on the telescope optics and is therefore
usually known, whereas the intensity of the source is a parameter to
be determined, we include the intensity but not the coherence length
in our parameter vector.

In Section \ref{pairsect}, we will consider a slightly more complicated
source, namely a pair of identical Gaussian--Schell sources. If the sources
are not correlated with each other, the coherence tensors can just be added,
hence the overall source coherence tensor has the form
\begin{eqnarray}\label{gauss-schell-pair}
Y(\bs{r}_1,\bs{r}_2) &=& \frac{I}{\sqrt{2\pi\sigma_\textrm{s}^2}}
\exp\left(-\frac{|\bs{r}_1-\bs{r}_2|^2}{2\sigma_\textrm{g}^2}\right)
\nonumber \\ &&
\times
\left(
\exp\left(-\frac{
|\bs{r}_1-\bs{r}_\textrm{c}-\bs{r}_\textrm{sep}/2|^2
+ 
|\bs{r}_2-\bs{r}_\textrm{c}-\bs{r}_\textrm{sep}/2|^2
}{4\sigma_\textrm{s}^2}\right)
+
\exp\left(-\frac{
|\bs{r}_1-\bs{r}_\textrm{c}+\bs{r}_\textrm{sep}/2|^2 
+ 
|\bs{r}_2-\bs{r}_\textrm{c}+\bs{r}_\textrm{sep}/2|^2
}{4\sigma_\textrm{s}^2}
\right)
\right) 
 \;,
\end{eqnarray}
where $\bs{r}_\textrm{c} \equiv (x_\textrm{c},y_\textrm{c})$ is the
position of the midpoint between the sources.
$\bs{r}_\textrm{sep} \equiv (x_\textrm{sep},y_\textrm{sep})$ is the 
separation vector between the sources. We define the parameter vector for
the double Gaussian--Schell source
as
\begin{eqnarray}\label{paramvectorpair}
\bs{\theta} = (x_\textrm{c},y_\textrm{c},x_\textrm{sep},y_\textrm{sep},I,\sigma_\textrm{s})^\mathrm{T} \;.
\end{eqnarray}  
In order that the posterior is unimodal (and hence that the Gaussian 
approximation can be applied), we impose the requirement that
$x_\textrm{sep} \geq 0$, which removes the inherent transpositional degneracy
of the problem.

\subsection{Parameterization of bolometric detectors}
In general the coherence tensor of a detector will depend on the details
of the physics. For the purpose of this paper however we require
a general parametrization that is a reasonable approximation to a typical
detector, and that incorporates the most important features.
Therefore we consider square pixels
of uniform absorbtivity, and with a Gaussian correlation function.
Such a model is partially coherent and hence multimode.
Thus the detector coherence tensor of the $a$th detector
centred at $(x^a,y^a)$ is given by
\begin{eqnarray}\label{detector}
Z^a_{ij}(\bs{r}_1,\bs{r}_2) &=&  
\Theta(x_1-(x^a-s/2))\Theta((x^a+s/2)-x_1) 
\Theta(x_2-(x^a-s/2))\Theta((x^a+s/2)-x_2) \nonumber \\ &&
\Theta(y_1-(y^a-s/2))\Theta((y^a+s/2)-y_1) 
\Theta(y_2-(y^a-s/2))\Theta((y^a+s/2)-y_2) 
\exp\left(-\frac{|\bs{r}_1-\bs{r}_2|^2}{2c}\right) \;,
\end{eqnarray}
where $s$ is the detector size, $c$ is the detector coherence length,
$\bs{r}_n = (x_n,y_n)$, and $\Theta(x)$ is the Heaviside step function.

\begin{figure}
\begin{center}
\makebox[3mm][l]{\hspace{7mm}(a)}
\includegraphics[angle=-90,width=0.3\linewidth]{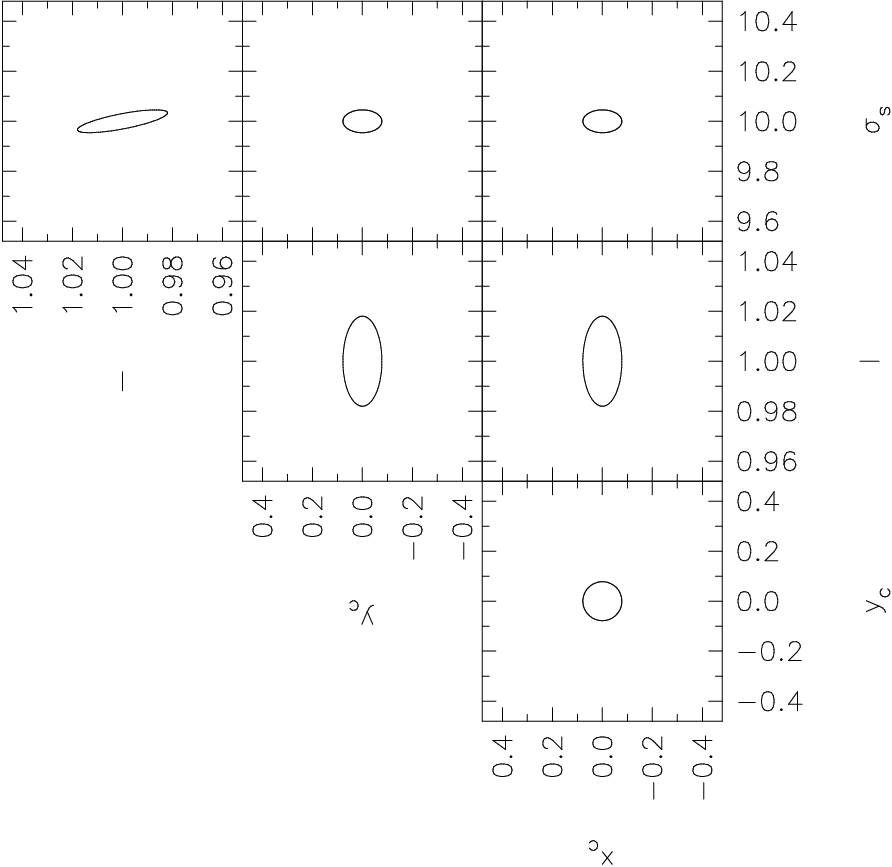}
\makebox[3mm][l]{\hspace{7mm}(b)}
\includegraphics[angle=-90,width=0.3\linewidth]{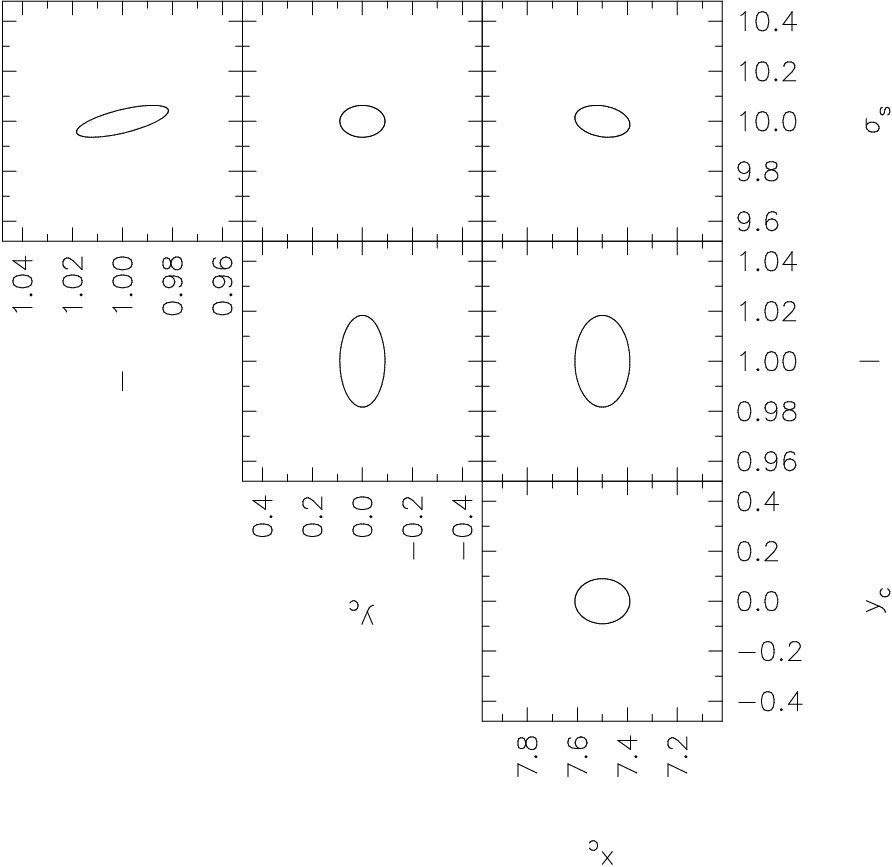}
\makebox[3mm][l]{\hspace{7mm}(c)}
\includegraphics[angle=-90,width=0.3\linewidth]{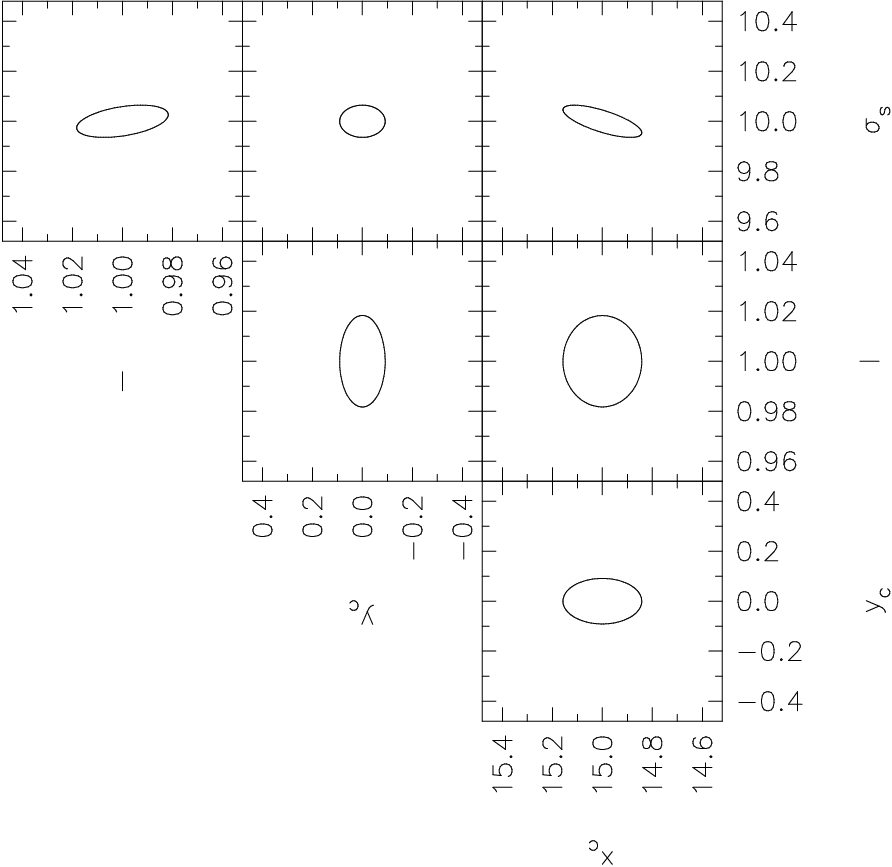}
\par
\makebox[3mm][l]{\hspace{7mm}(d)}
\includegraphics[angle=-90,width=0.3\linewidth]{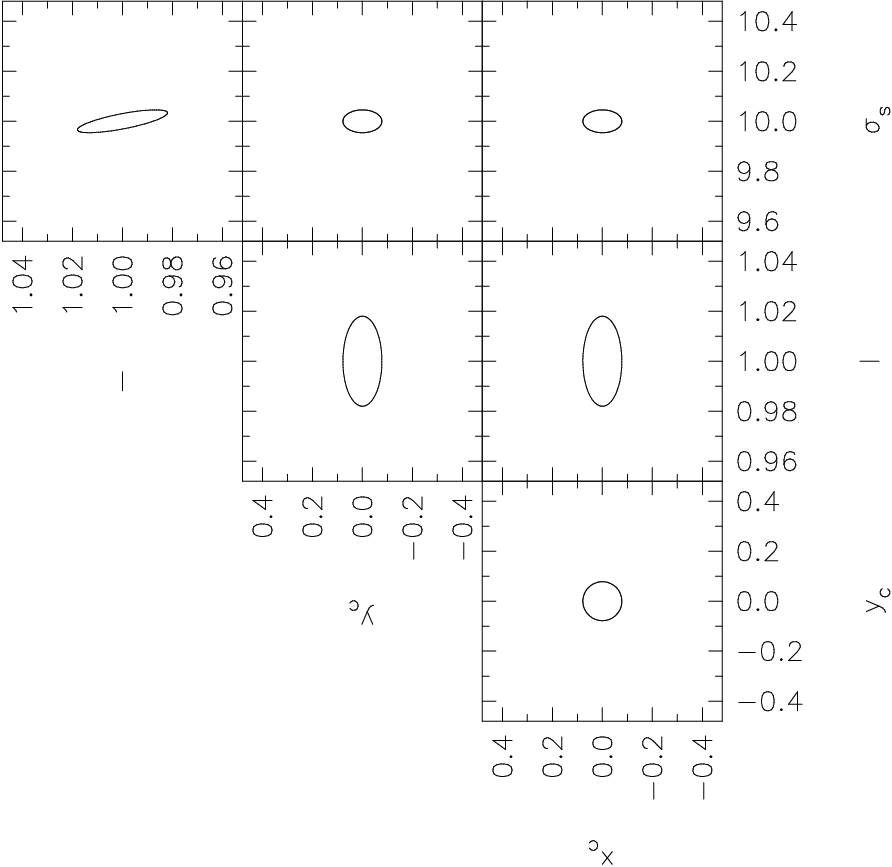}
\makebox[3mm][l]{\hspace{7mm}(e)}
\includegraphics[angle=-90,width=0.3\linewidth]{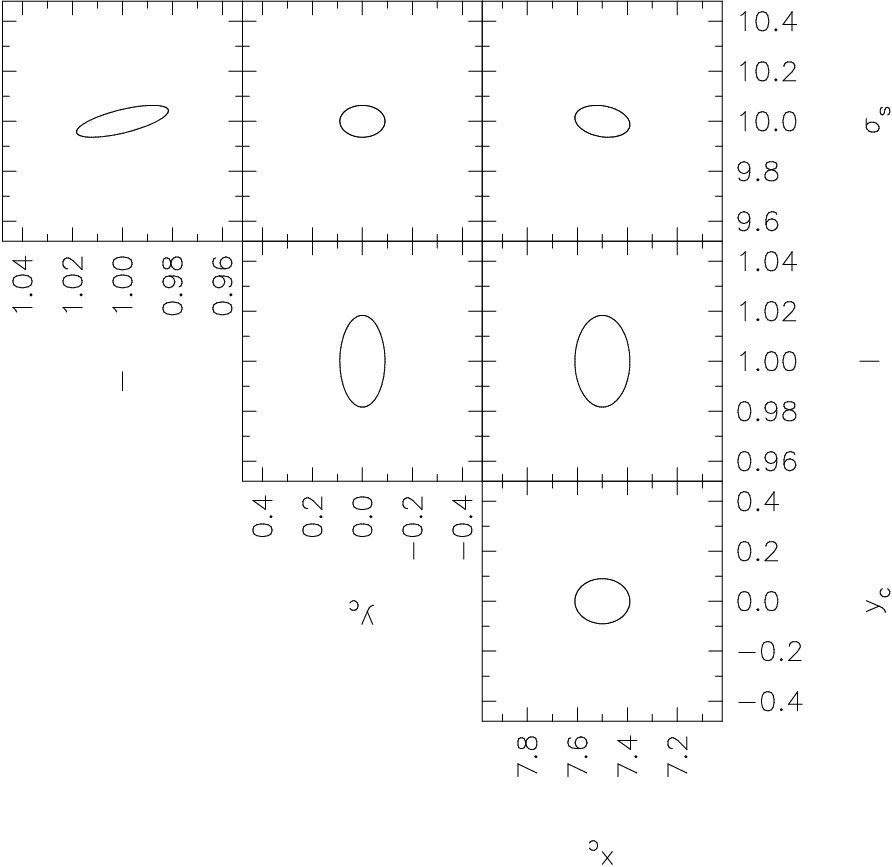}
\makebox[3mm][l]{\hspace{7mm}(f)}
\includegraphics[angle=-90,width=0.3\linewidth]{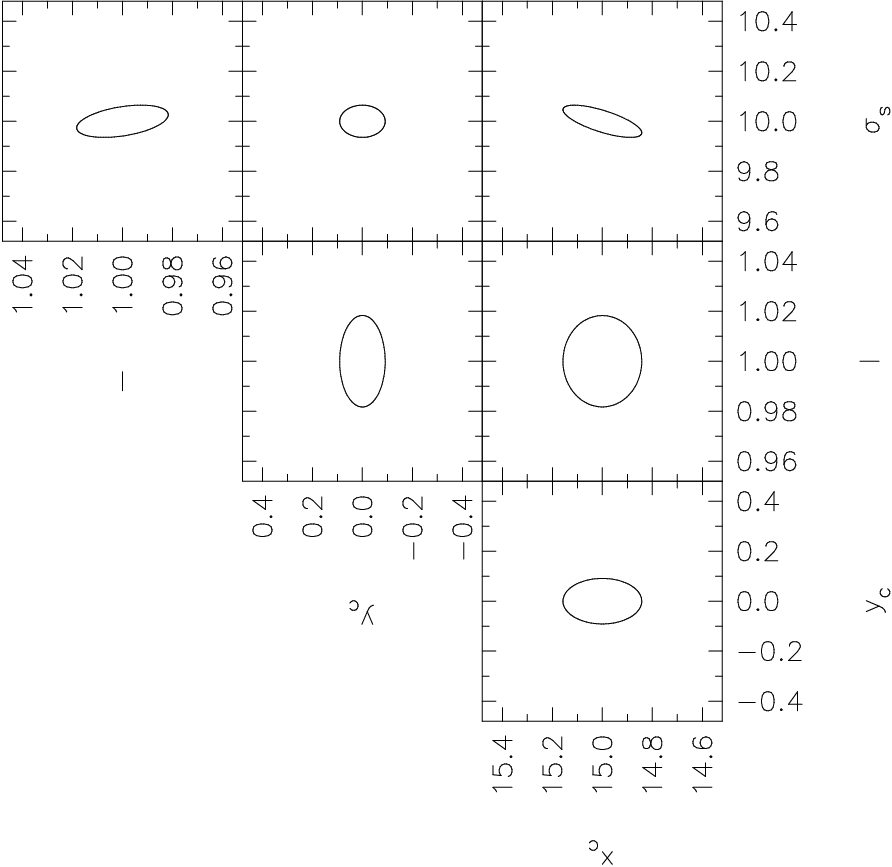}
\caption{\label{fig1}
Estimated 80\% marginalized confidence contours
for the recovered parameters of a Gaussian--Schell source, observed with
$5\times5$ arrays of detectors centered 15 units apart. In all cases
the source has parameters $x_c=0$, $I=1$, $\sigma_s=10$ and $\sigma_g=5$,
while the detectors have coherence length $c=5$, and integration time
$\tau=100/\Delta\omega$.
On the top row (a--c), the detector are of side length 15 units, 
and are thus close packed,
while on the bottom row (d--f) they are of side length 3 units, 
and are therefore sparse.
The source is at the centre of the array ($x_c=0$) 
in the plots on the left (a, d), while the
plots in the middle (c, e) and on the right (d, f) have 
$x_c=7.5$ and $x_c=15$ respectively.
}
\end{center}
\end{figure}

\section{Simulations}
We perform all simulations on a uniform $51 \times 51$ grid. 
We use (\ref{zdisc})
and (\ref{ydisc}) to perform the discretization, (\ref{datavector}),
(\ref{covmatrix}), (\ref{power}) and (\ref{narrowbandcov}) to calculate
the data vector and covariance matrix, and (\ref{fisherapprox}) to
calculate the Fisher matrix. In the parametrizations for source and
detector that follow, we measure all lengths in number of grid points;
our results can straightforwardly be scaled to physical units.
We set the ratio of integration time to coherence time to be 100. 
Numerical differencing with a step size of 0.001 was used 
to calculate the derivatives $\partial D_k / \partial \theta_i$.
We verified, by using polynomial interpolation techniques to
calculate the derivatives to machine precision in a few random cases,
that numerical differencing gives more than sufficient accuracy
for the purpose, as well as being much faster.
Inversion of the covariance matrix is problematic due to its poor
conditioning. To minimise numerical errors and
make the inversion stable we therefore employ the pseudo--inverse
\citep{moore1920:1,penrose1955:1}, discarding singular values below one 
millionth of the
largest singular value. This cut--off is arbitrary, but we found
our results were stable through several orders of magnitude of cut--off.
Physically, we can justify the use of the pseudo--inverse from the fact that
the signal--to--noise ratio falls off rapidly at the edge of the source
\citep*{saklatvala2006:4}, 
and hence these detectors do not contribute significantly to the Fisher
matrix. The effect of taking the pseudo--inverse of the covariance
matrix is to discard these same detectors; hence the result is
almost identical to that which would be obtained using the true inverse
on a machine of unlimited precision.  

\subsection{Parameter recovery for a Gaussian--Schell source \label{gs1}}
Figure \ref{fig1} shows the $80\%$ marginalized confidence contours for
the recovered parameters of a Gaussian--Schell source, estimated using
the Fisher matrix, when the source is observed with
close packed and sparse arrays (see figure caption for details).
We observe that all parameters are better constrained with the
close packed array, as expected. In all the plots shown,
the width of the source is more tightly constrained than its position; this
can be predicted from the fact that the source is constrained in the
model to be circularly symmetric, and thus changing $\sigma_s$ affects the
extent of the source in both $x$ and $y$ directions, and thus affects the
output of more detectors than a change in the $x$ or $y$ coordinate
alone. When the source is at the centre of the array, the parameters are 
uncorrelated
except the width and intensity, which are anti--correlated.
In some of our subsequent simulations the width and intensity
are positively correlated. 
A common theme of this paper is that while
our simulations produce all the results that could be straightforwardly
predicted, they also uncover many other effects that might
not have been expected.

Figure \ref{fig1} also shows the effect of moving a source away from the
centre. We choose the displacements so that the source is centered
midway between detectors (centre), and centered on a detector adjacent
to the central one (right). We note that there is a minimal 
difference when the source is centred between detectors, even for the sparse
array, where the gaps are 12 units, and thus greater than $\sigma_s$. 
This suggests that as long as neither the gaps between the detectors
nor the detector spacing
are much larger than $\sigma_s$, the position of the source does
not affect the amount
of information that can be recovered. The implication is that
the shoulders of the Gaussian intensity distribution are the important
features,
not the central peak. Our simulations show on the other hand that losing part
of the source off the edge of the array has a dramatic effect on the estimate
of the displacement, though less so on the other parameters. It also
introduces a correlation between the displacement and the width. When
the source is centered on a detector adjacent to the central one,
it is centered 
roughly $1\sigma_s$ from the edge of the array, and these effects are
seen clearly.

\begin{figure}
\begin{center}
\makebox[3mm][l]{\hspace{3mm}(a)}
\includegraphics[angle=-90,width=0.7\linewidth]{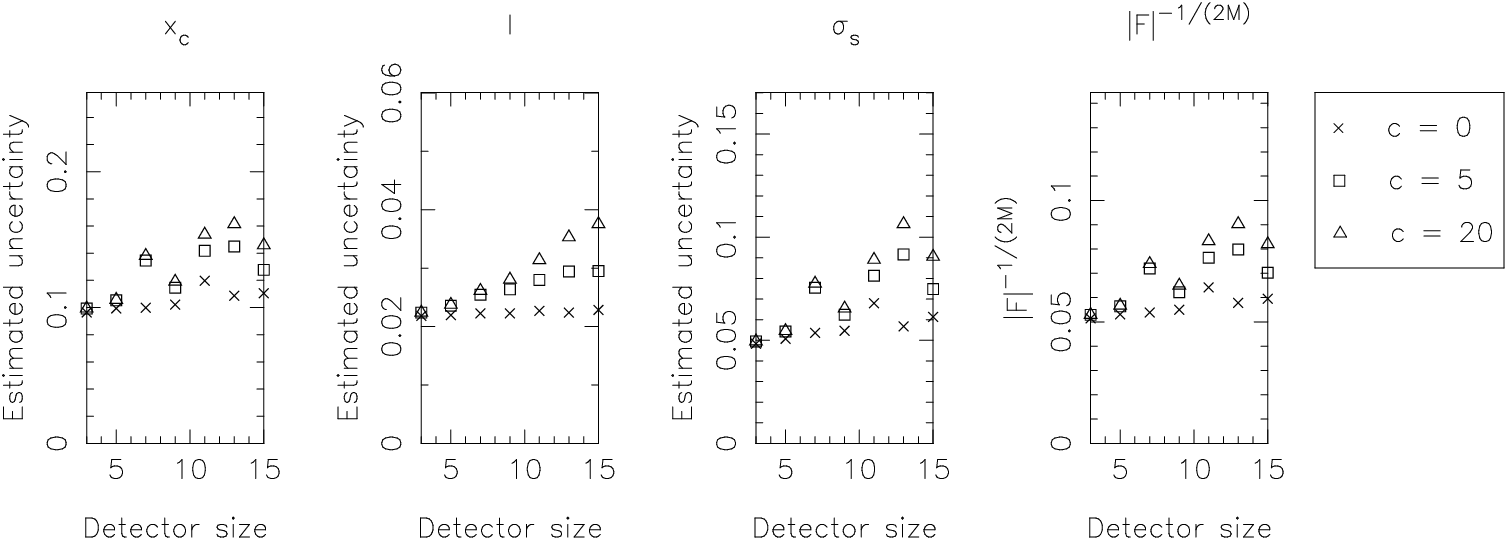}
\par \vspace{2mm}
\makebox[3mm][l]{\hspace{3mm}(b)}
\includegraphics[angle=-90,width=0.7\linewidth]{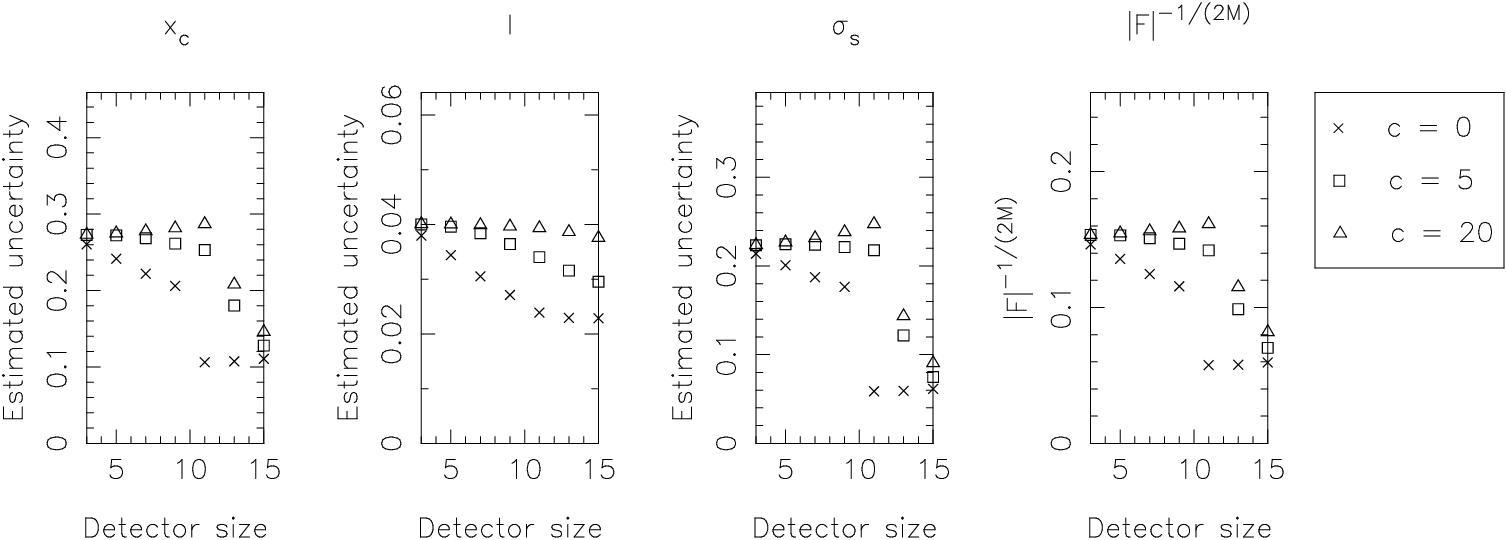}
\caption{\label{fig2}
Estimated uncertainties on the recovered parameters ($x_c$ and $y_c$ have
the same uncertainties), and the $M$th root (where
$M$ is the number of free parameters in the model, equal to 4), 
of the estimated allowed volume of parameter space, for a Gaussian--Schell 
source with parameters $x_c=y_c=0$, $I=1$, $\sigma_s=10$ and $\sigma_g=5$,
observed with arrays of detectors of various size and coherence length (c).
In the first row of plots (a), the detectors are close packed,
with one detector centered at $(0,0)$, 
and a minimum total size of $51\times 51$
units. In the second row (b), the array contains $5\times 5$ detectors centered
15 units apart, such that the array becomes increasingly sparse as the
detector size is reduced.
}
\end{center}
\end{figure}

\begin{figure}
\begin{center}
\makebox[3mm][l]{\hspace{3mm}(a)}
\includegraphics[angle=-90,width=0.7\linewidth]{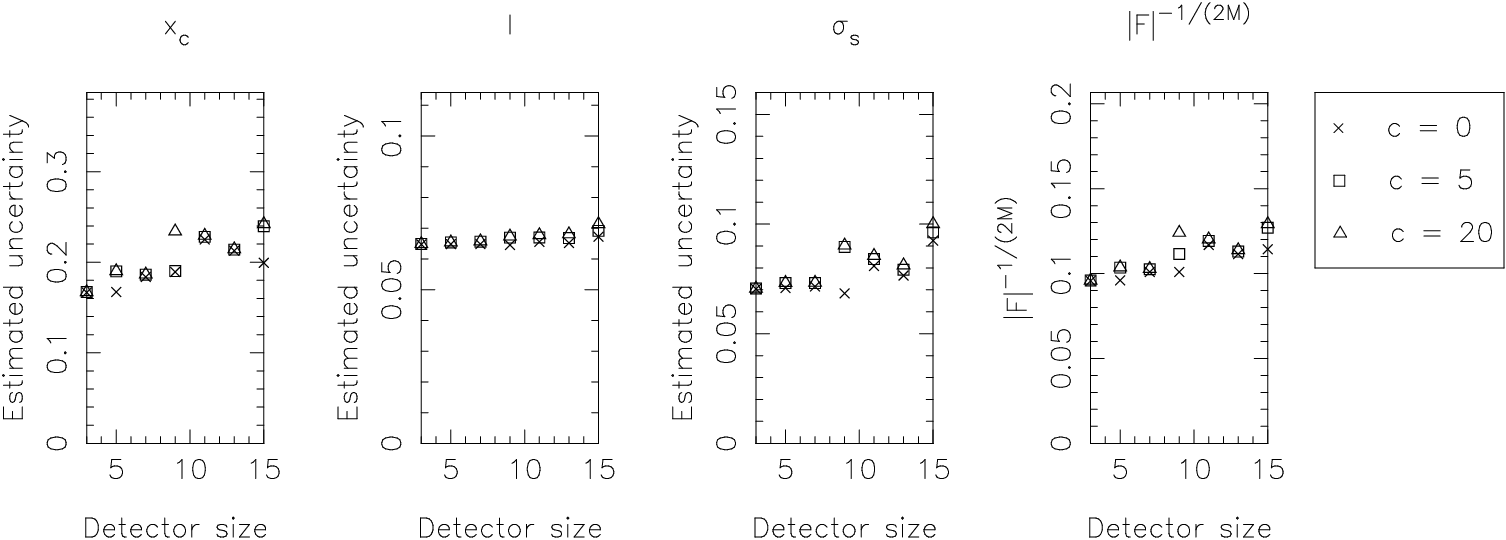}
\par \vspace{2mm}
\makebox[3mm][l]{\hspace{3mm}(b)}
\includegraphics[angle=-90,width=0.7\linewidth]{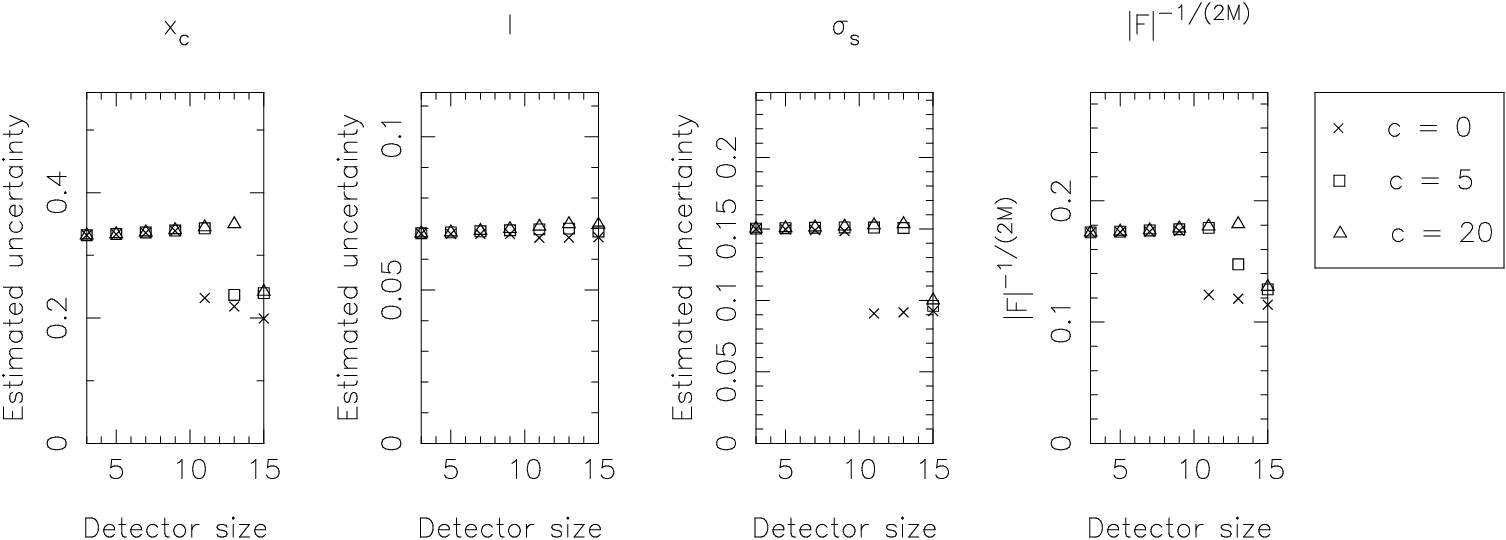}
\caption{\label{fig3}
Estimated uncertainties on the recovered parameters, and the $M$th root
of the estimated allowed volume of parameter space, for a Gaussian--Schell 
source observed with close packed and sparse arrays of detectors of 
various size and coherence length.
The sources and arrays are identical to those in Figure \ref{fig2},
except that the coherence length of the source is increased such that
$\sigma_g = 20$.
}
\end{center}
\end{figure}

\begin{figure}
\begin{center}
\makebox[3mm][l]{\hspace{3mm}(a)}
\includegraphics[angle=-90,width=0.7\linewidth]{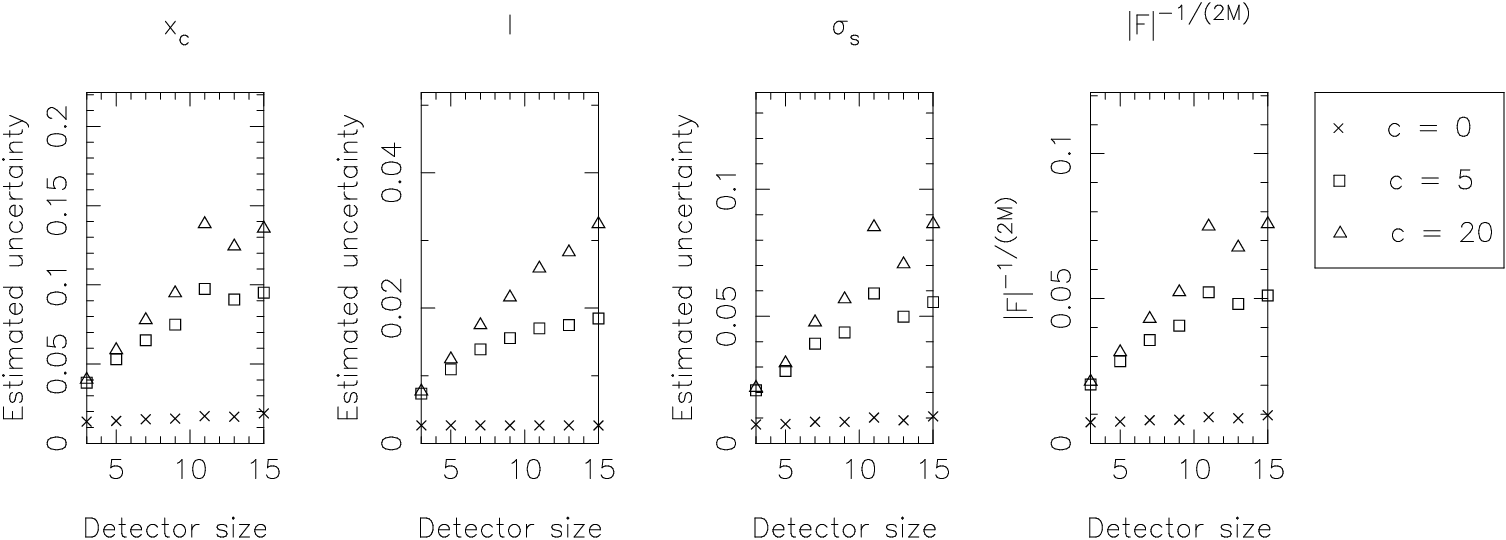} 
\par \vspace{2mm}
\makebox[3mm][l]{\hspace{3mm}(b)}
\includegraphics[angle=-90,width=0.7\linewidth]{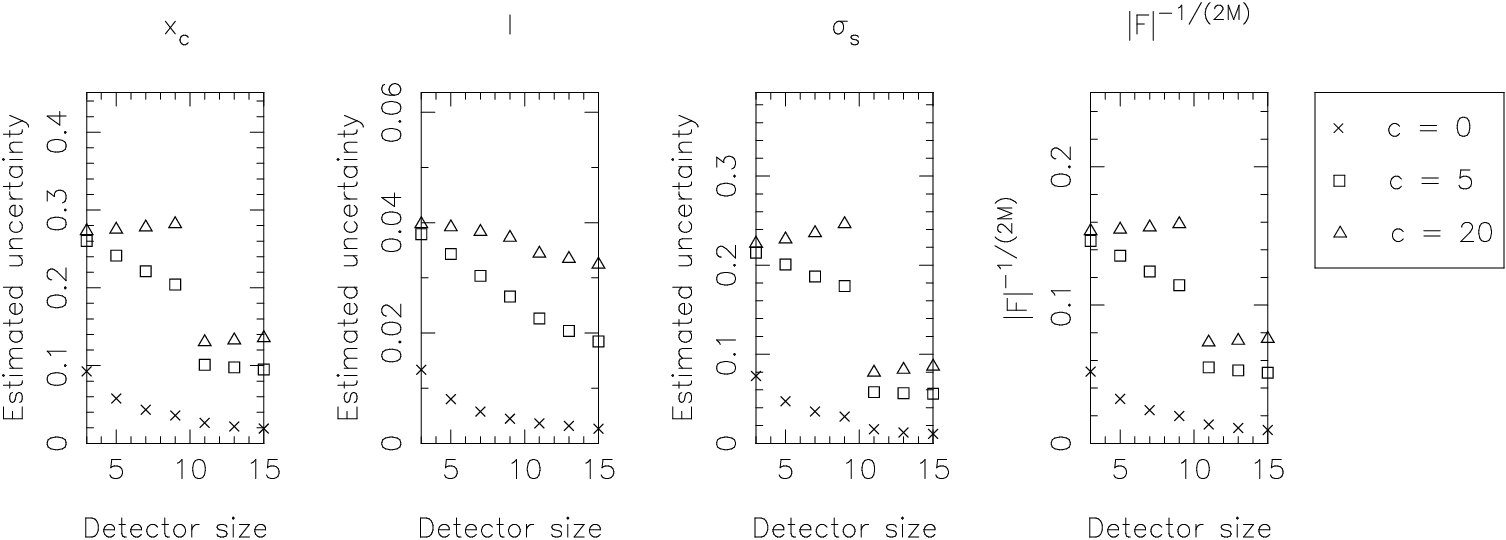}
\caption{\label{fig4}
Estimated uncertainties on the recovered parameters, and the $M$th root
of the estimated allowed volume of parameter space, for a Gaussian--Schell 
source observed with close packed and sparse arrays of detectors of 
various size and coherence length.
The sources and arrays are identical to those in Figure \ref{fig2},
except that the source is incoherent
($\sigma_g = 0$).
}
\end{center}
\end{figure}

\begin{figure}
\begin{center}
\makebox[3mm][l]{\hspace{3mm}(a)}
\includegraphics[angle=-90,width=0.7\linewidth]{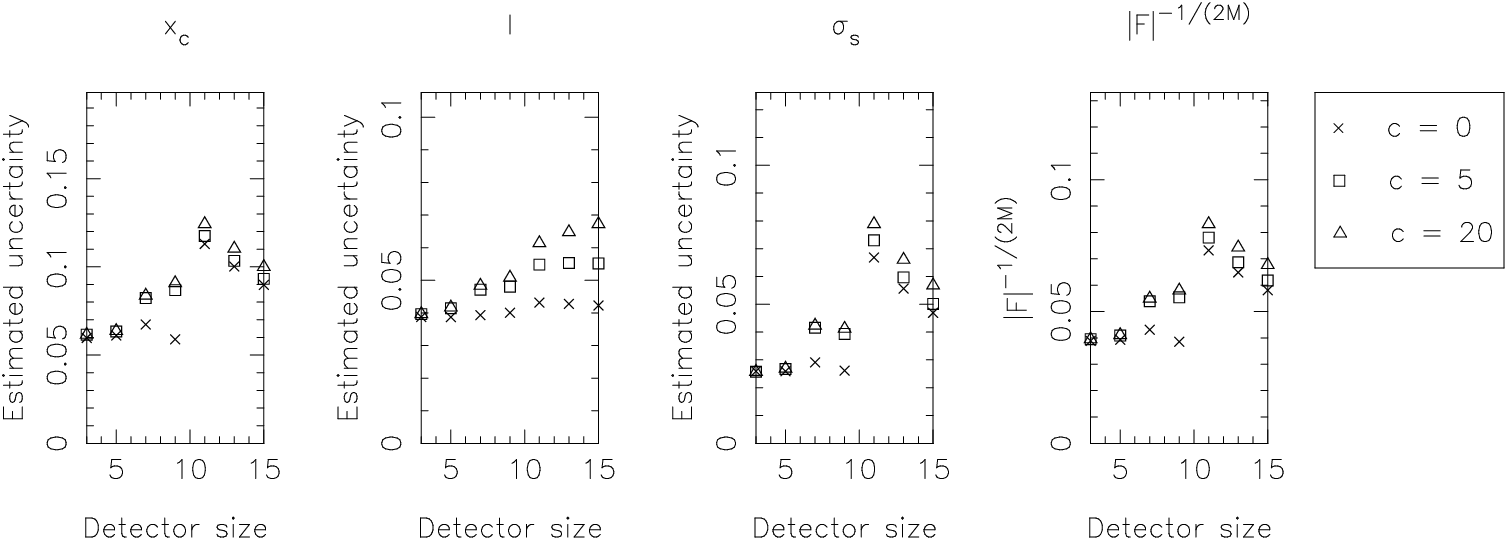}
\par \vspace{2mm}
\makebox[3mm][l]{\hspace{3mm}(b)}
\includegraphics[angle=-90,width=0.7\linewidth]{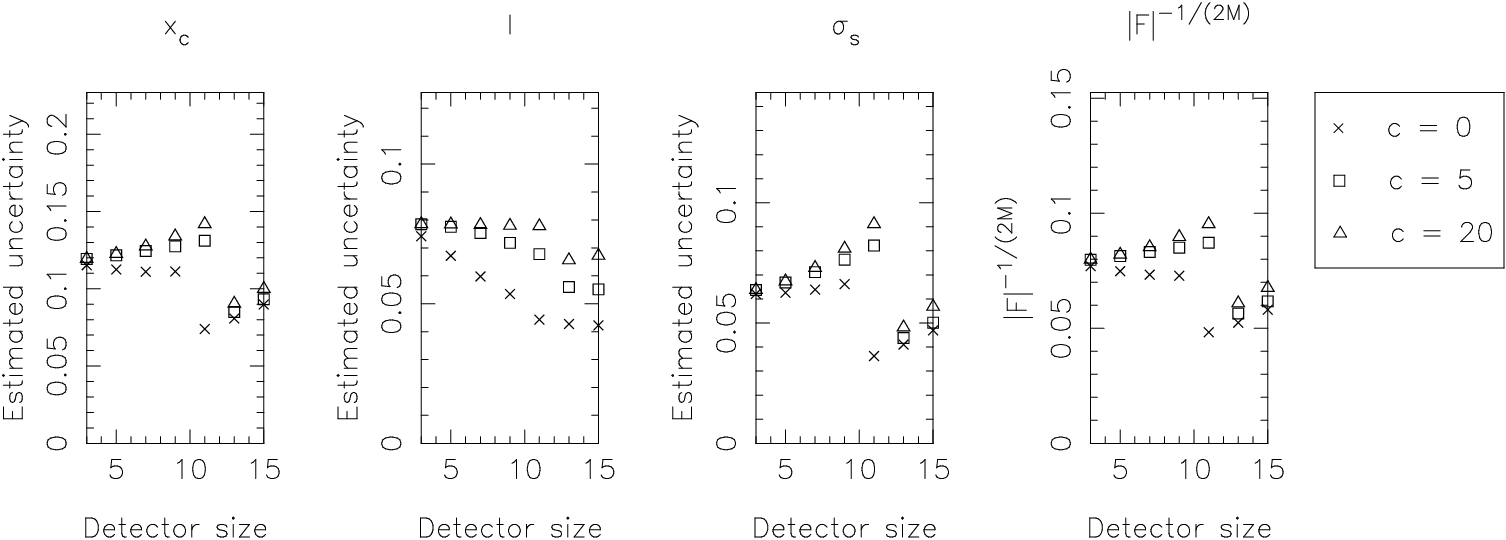}
\end{center}
\caption{\label{fig5}
Estimated uncertainties on the recovered parameters, and the $M$th root
of the estimated allowed volume of parameter space, for a Gaussian--Schell 
source observed with close packed and sparse arrays of detectors of 
various size and coherence length.
The sources and arrays are identical to those in Figure \ref{fig2},
except that the geometric size of the source is reduced such that
$\sigma_s = 5$.
}
\end{figure}

\subsection{Effect of detector size and coherence on close--packed and
sparse arrays}
We have seen that sparse arrays of small detectors are less effective
at recovering source parameters than close packed arrays of large detectors,
but now we investigate in detail the effects of both detector size
and coherence, and distinguish between the effect of detector size and the
effect of the number of detectors in the array.

Figure \ref{fig2} shows the estimated uncertainty on the recovered parameters
of the same Gaussian--Schell source as in Section \ref{gs1}, with close packed 
and sparse arrays of detectors of various size and coherence
(see figure caption for details). We also show the quantity
$|\bs{F}^{-\frac{1}{2M}}|$, where $\bs{F}$ is the Fisher matrix and $M$ is the 
number of parameters in the model. This quantity can be interpreted 
as the $M$th root
of the estimated allowed 
volume of parameter space, and gives a measure of the overall
uncertainty in the measurement. If the parameters are uncorrelated,
it will be equal to the geometric mean of the uncertainties on the individual
parameters, but if they are correlated it will be less. 

We can identify many trends. The more coherent the detector
the less accurately the source parameters can be determined,
as would be expected from our detector model, in which more coherent
detectors have inferior signal to noise. Furthermore, this effect
becomes more significant as the detector size increases; again this
is to be expected, as the coherence makes little difference to a small
detector. A final general comment
is that the uncertainty in the intensity follows smooth trends
for both close packed and sparse arrays, while the behaviour of the 
position parameters, which is more complicated, is always the same as that
of the width.

For the close packed arrays, the uncertainty
of the parameter estimates generally increases with detector size. However, 
the detailed behaviour is much more complicated. The incoherent
detectors give almost identical accuracy regardless of the size. The 
uncertainty on the intensity increases progressively with detector size
whereas the uncertainties on the position and width rises and falls
twice, but with an underlying increasing trend.
The tendency to less accurate parameter recovery with increasing size
can be attributed only in part to reduced spatial resolution,
as the effect is much less for incoherent detectors.
Rather the trends seen appear to be caused by the details of the coupling
of the natural modes of the detector to the natural modes of the source.

In the case of the sparse arrays with fixed number of pixels, we find in all 
cases that the uncertainty in the intensity falls gradually
as the detector size increases, while there is a step change in the
other parameters at a size of around 11 units. The detailed
behaviour depends on the coherence length; for the coherent detectors,
the uncertainty actually \emph{increases} with size up to a threshold, then
falls, whereas for the less coherent detectors the uncertainty falls gradually
up to the threshold, then suddenly, before plateauing. Furthermore,
the precise threshold size depends on the coherence length. 

We now investigate how the size and coherence length of the source
changes our results. Figure \ref{fig3} shows the same information as Figure
\ref{fig2}, except that the coherence length of the source
is increased fourfold. We see a number of changes in the behaviour. The
effect of the coherence of the detector is now generally much less pronounced. 
The uncertainty
on the intensity is almost independent of detector size. The uncertainty
on the position and width for the close packed arrays 
show the same complicated behaviour as before,
but the peaks are shifted. For the sparse arrays, we still see the step change
at around 11 units, but almost no variation below that threshold. The
coherence length of the detector does affect the position of this
transition, hence when the size is around 11 units the detector 
coherence length \emph{does} make a 
significant difference. Finally we see that the uncertainties on
all parameters are greater than for the less coherent source, as expected,
because the signal to noise is worse.

Figure \ref{fig4} shows the same information as Figure
\ref{fig2}, except that the source
is incoherent. The opposite effects are seen as in Figure \ref{fig3}:
the detector coherence length is much more important, the step change for
the sparse arrays is still present, though less obviously for
the incoherent detector, whereas the other trends are more pronounced.
The peaks and troughs for the position and uncertainty with the 
close packed arrays are shifted yet again, though not predictably.
The uncertainties on
all parameters are less than for the partially coherent sources, as expected.

Figure \ref{fig5} shows the same information as Figure
\ref{fig2}, except that the size of the source
is halved. 
Most of the trends are very similar, but the obvious difference
is the sharp peak in the uncertainty of the position and width
estimates for the close packed array with size 11 units. There is also some 
slight discrepancy in behaviour between the position and width estimates,
which was not seen in Figure \ref{fig2}. More telling however is the fact
that the step change for the sparse arrays is not shifted either
by changing the size or coherence of the source. It is not therefore simply
caused by matching the size of the detectors to either size or
coherence length of the source; rather the most important factor
seems to be the fraction of the focal plane covered by the detectors.
Finally, the errors on the position and size are less than in Figure 
\ref{fig2}, though the error on the intensity is greater.

In this section, we have seen that apart from a few predictable trends,
most of the behaviour of imaging arrays is quite complicated and
not possible to predict by straightforward physical arguments. The
complexity appears to be due to the details of how the natural modes of the
source (in this case the Gauss-Hermite modes) couple to the modes
of the detectors. Indeed, 
when the length scales of source and detectors
are similar, and the sources and detectors are partially coherent,
there is no reason to expect simple behaviour.
Our simulations have demonstrated the importance of performing detailed
simulations whenever partially coherent arrays are to be designed.
Nevertheless, there are a number of key trends to emerge.
Firstly the coherence of both detector and source have a significant effect
on the parameter recovery. Secondly, the packing density needs to be above a 
certain threshold for information to be recovered. This packing density is 
roughly independent of the source parameters, and corresponds to roughly
$50\%$ of the focal plane being filled.

\begin{figure}
\begin{center}
\makebox[3mm][l]{\hspace{7mm}(a)}
\includegraphics[angle=-90,width=0.3\linewidth]{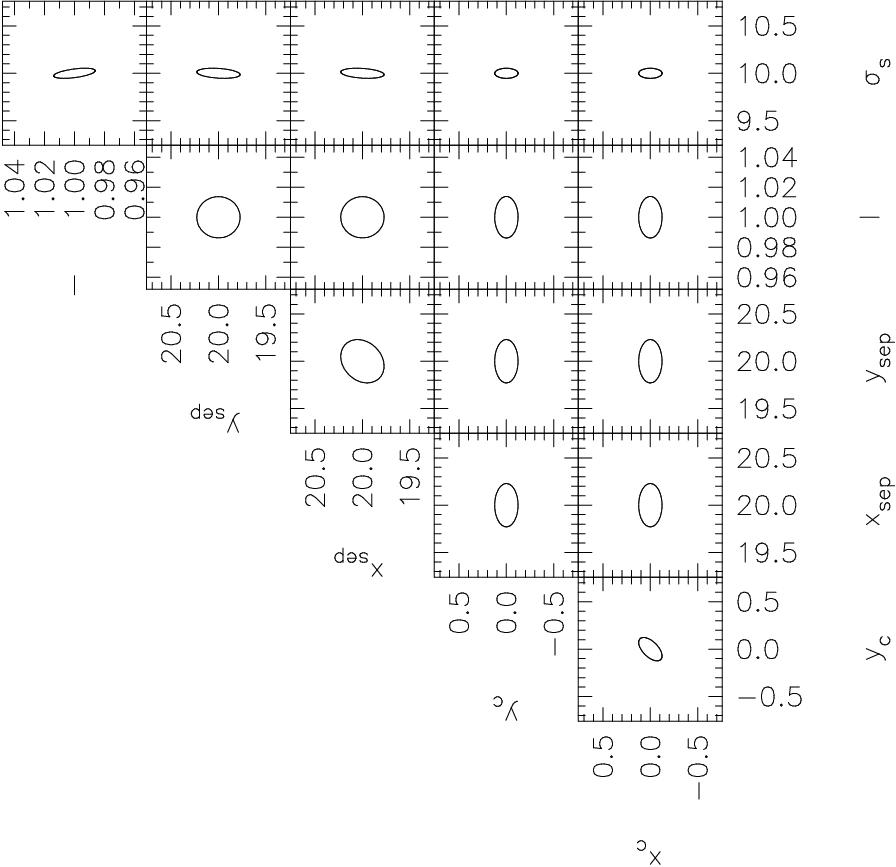}
\makebox[3mm][l]{\hspace{7mm}(b)}
\includegraphics[angle=-90,width=0.3\linewidth]{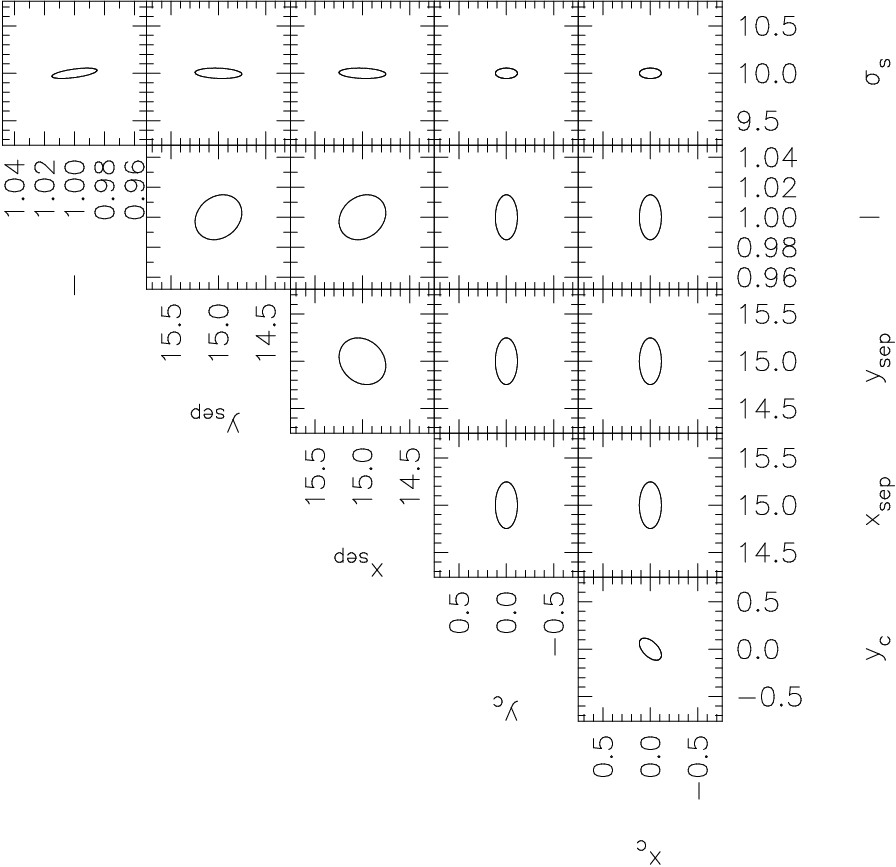}
\makebox[3mm][l]{\hspace{7mm}(c)}
\includegraphics[angle=-90,width=0.3\linewidth]{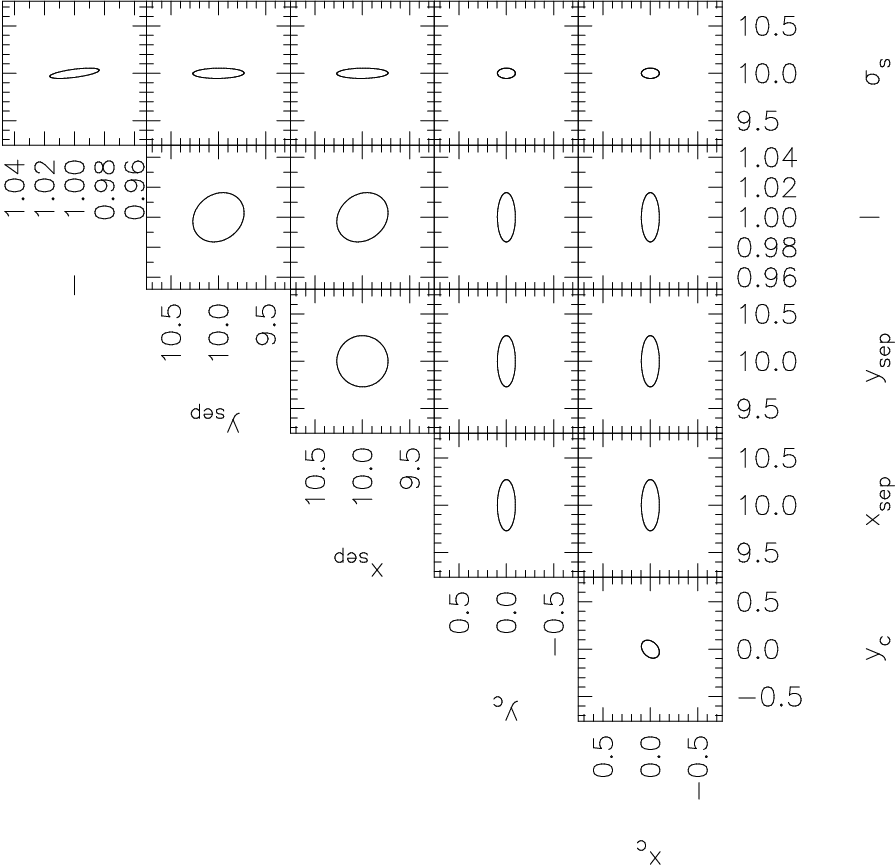}
\par
\makebox[3mm][l]{\hspace{7mm}(d)}
\includegraphics[angle=-90,width=0.3\linewidth]{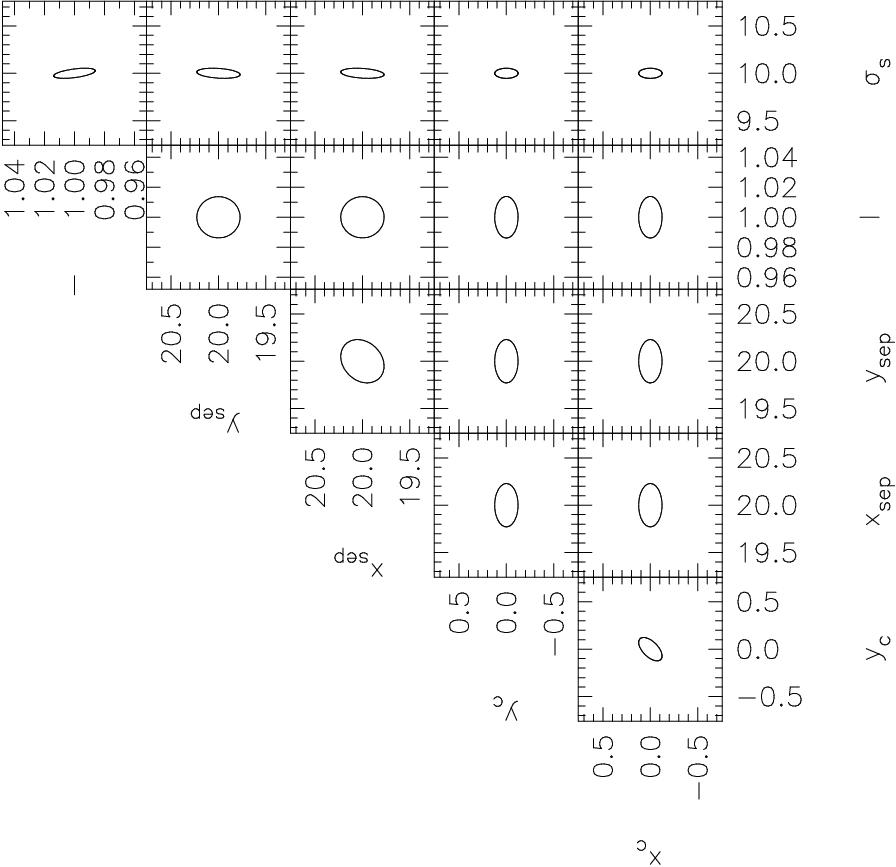}
\makebox[3mm][l]{\hspace{7mm}(e)}
\includegraphics[angle=-90,width=0.3\linewidth]{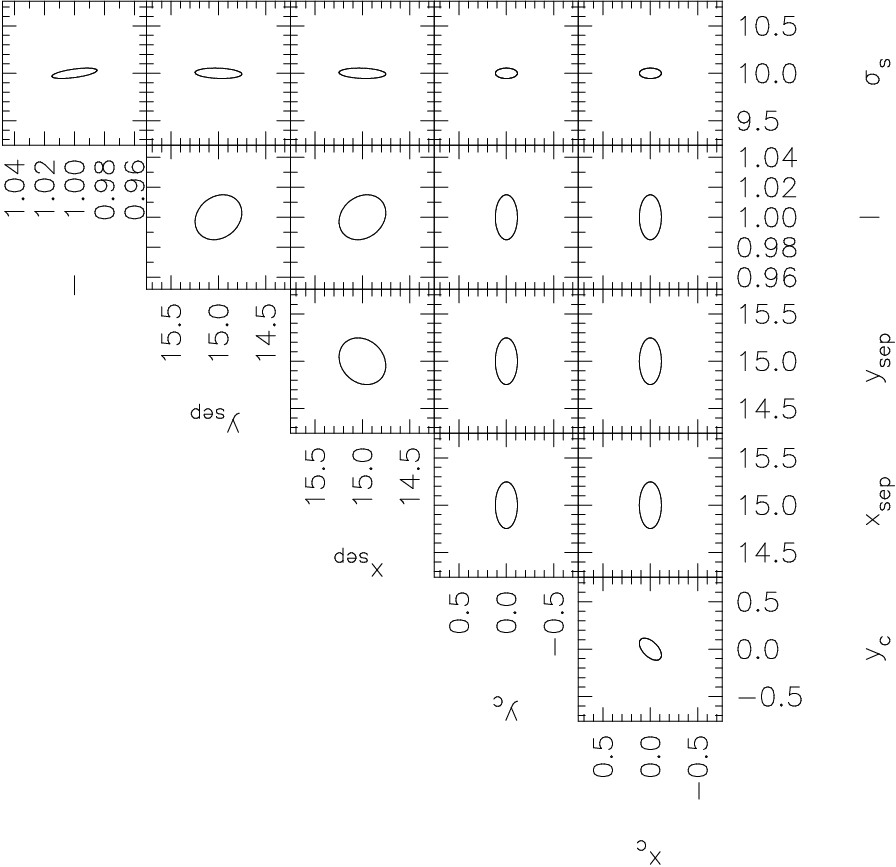}
\makebox[3mm][l]{\hspace{7mm}(f)}
\includegraphics[angle=-90,width=0.3\linewidth]{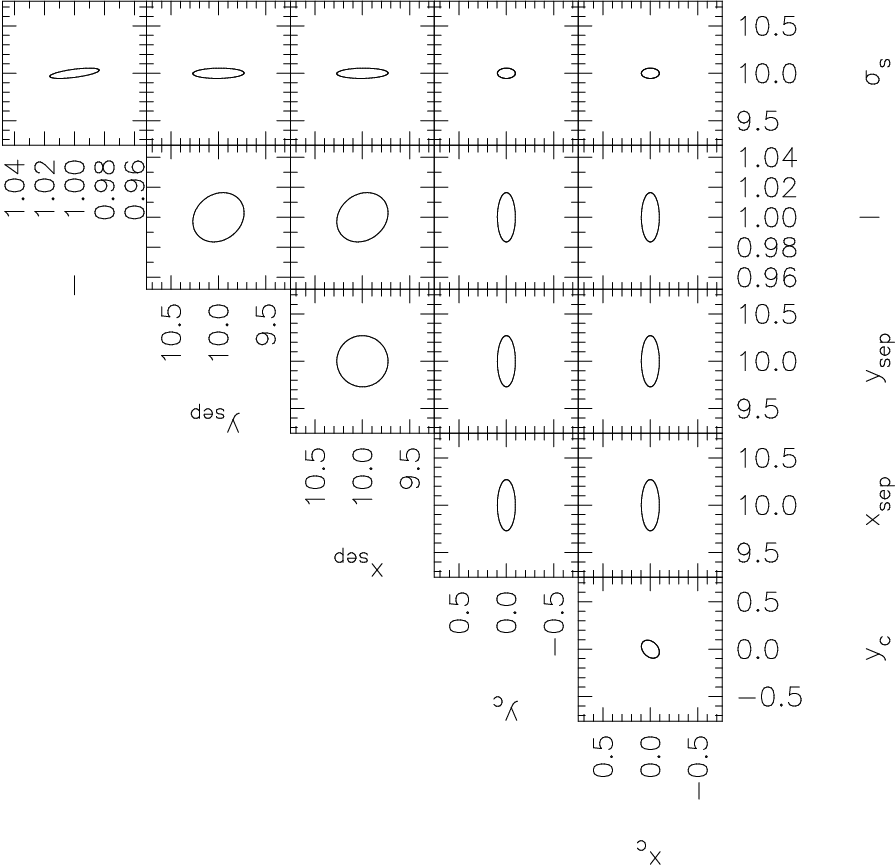}
\end{center}
\caption{\label{fig6}
Estimated 80\% marginalized confidence contours
for the recovered parameters of a pair of identical Gaussian--Schell sources, 
observed with $5\times5$ arrays of detectors centered 15 units apart.
In all cases
the source has parameters $x_c=y_c=0$, $y_{sep}=0$, $I=1$, $\sigma_s=10$ and 
$\sigma_g=5$,
while the detectors have coherence length $c=5$.
On the top row (a--c), the detector are of side 15 units, 
and are thus close packed,
while on the bottom row (d--f)
they are of side 3 units, and are therefore sparse.
The sources are separated in both the $x$ and $y$ directions 
with $x_{sep}=y_{sep}=10$ (a, d),
$x_{sep}=y_{sep}=15$ (b, e) and $x_{sep}=y_{sep}=20$ (c, f).
}
\end{figure}

\subsection{Parameter recovery for a double Gaussian--Schell source \label{pairsect}}
We now consider a slightly more complicated source: the double
Gaussian--Schell source defined by (\ref{gauss-schell-pair}).
Figure \ref{fig6} shows the 80\% confidence contours of the
parameters of such sources with various separation
vectors, when observed with both
close--packed and sparse arrays. As the sources are moved apart in the
$x$ direction, $x_c$ becomes better constrained, but the other
parameters become less well constrained, due to part of the source moving 
off the edge of the array. The same effect is seen for both close packed and 
sparse arrays.
The source parameters can be equally well inferred even when the 
sources are strongly overlapping. However, this is not to say that
the sources can be resolved: in order to determine whether two
sources are resolved, it is necessary to compare models,
for instance a pair of identical circularly symmetric sources with
a single elliptical source. Fisher matrix analysis is not
sufficient for this purpose; however, our coupled mode theory
of detectors allows the higher moments,
and hence the full likelihood function of the detector outputs to be 
calculated, and thus, in principle, 
it enables a detailed Bayesian analysis to be carried out.
For the purpose of this paper, we simply consider Figure \ref{fig6} to be
a demonstration of how the Fisher matrix technique 
can be applied to complex sources
of known form.

\begin{figure}
\begin{center}
\makebox[3mm][l]{\hspace{7mm}(a)}
\includegraphics[angle=-90,width=0.3\linewidth]{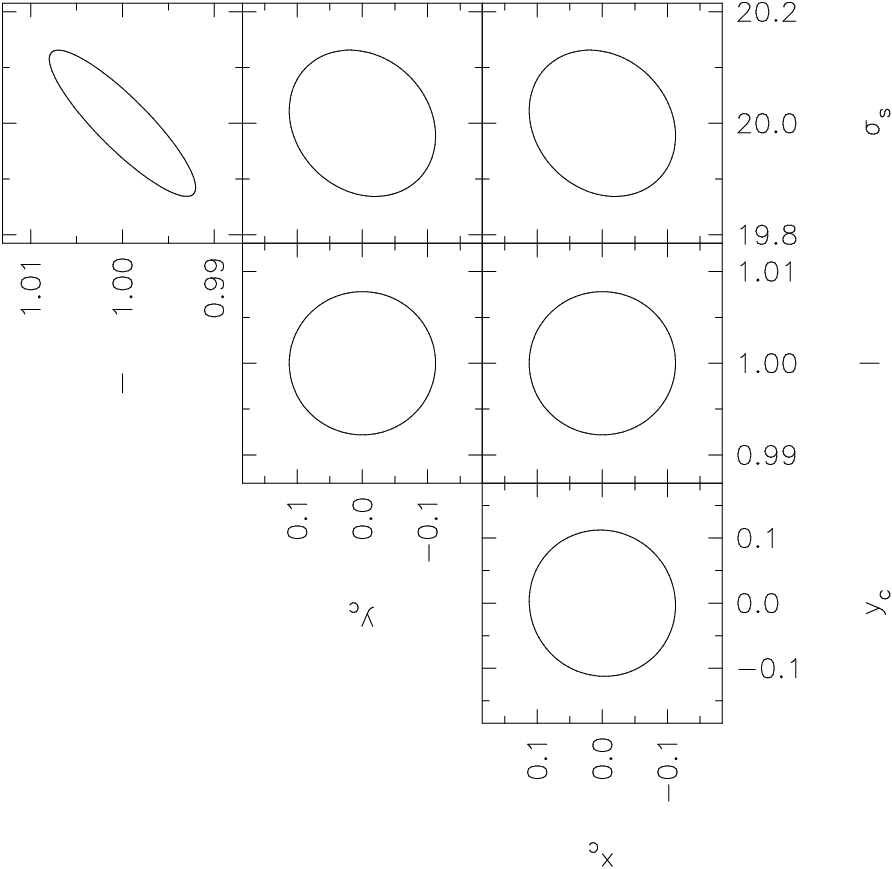}
\makebox[3mm][l]{\hspace{7mm}(b)}
\includegraphics[angle=-90,width=0.3\linewidth]{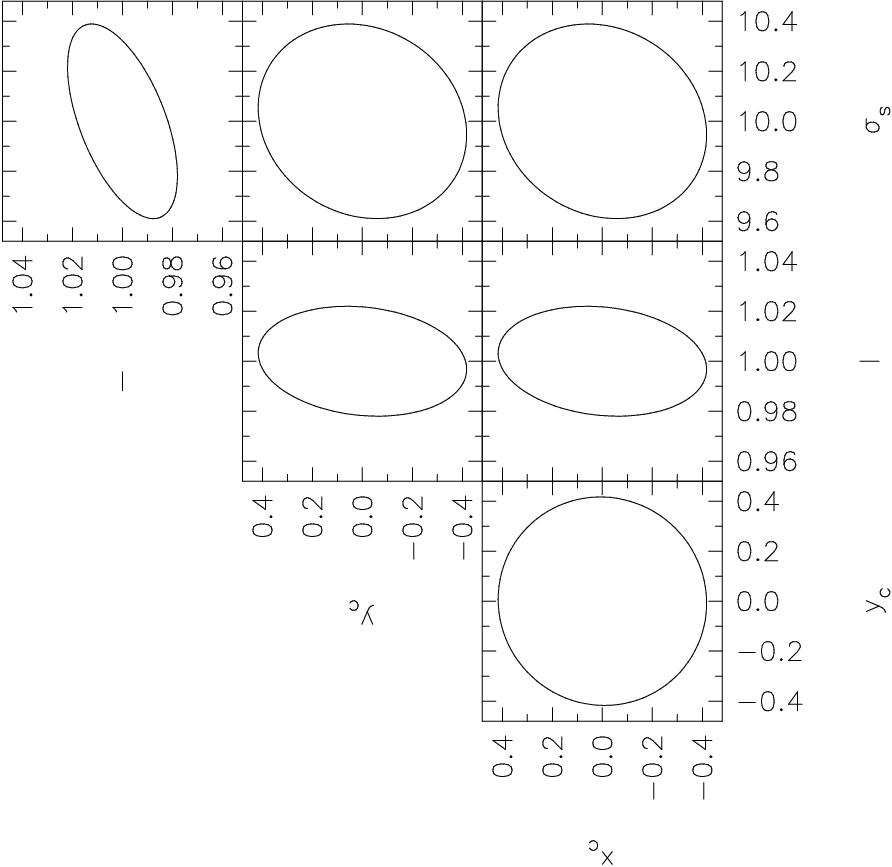}
\makebox[3mm][l]{\hspace{7mm}(c)}
\includegraphics[angle=-90,width=0.3\linewidth]{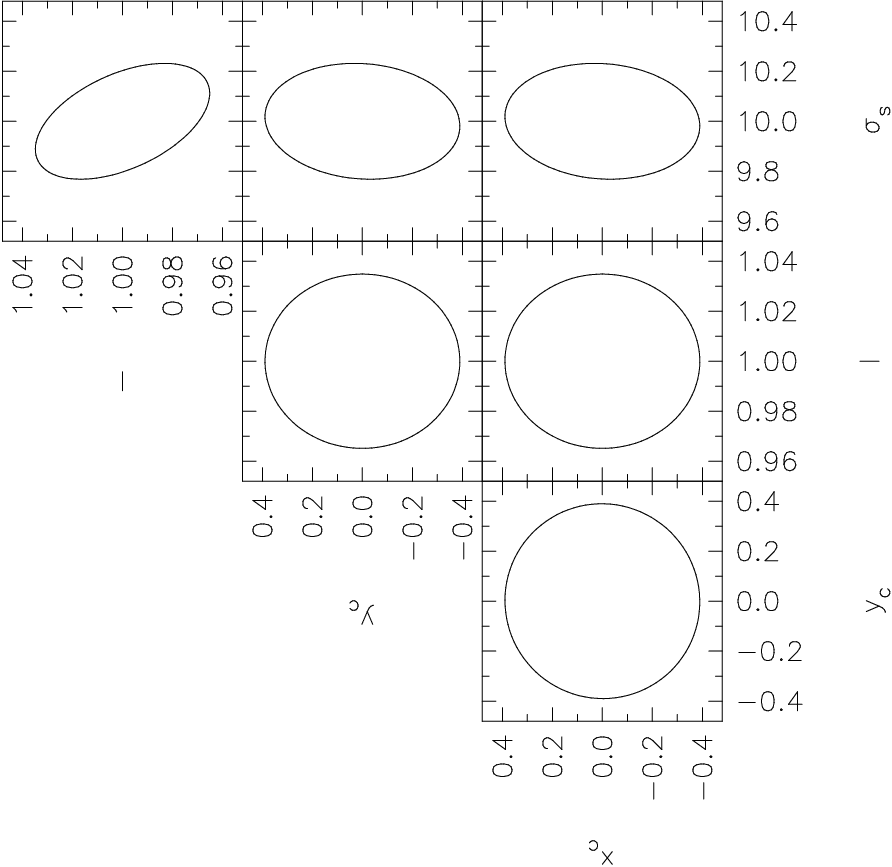}
\par
\makebox[3mm][l]{\hspace{7mm}(d)}
\includegraphics[angle=-90,width=0.3\linewidth]{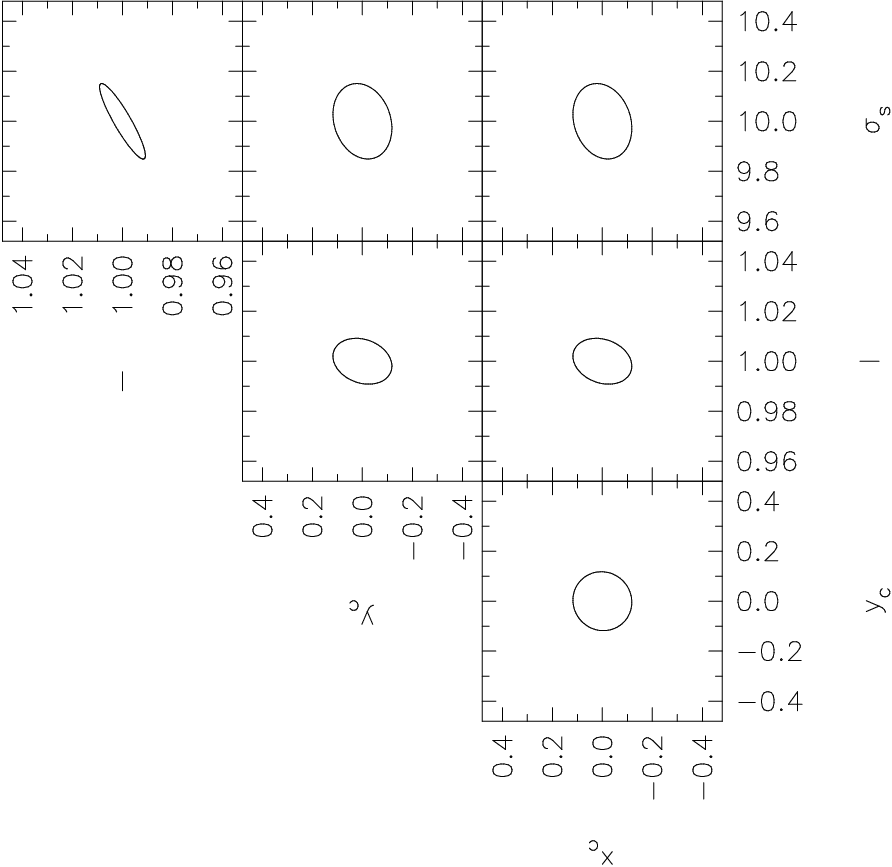}
\makebox[3mm][l]{\hspace{7mm}(e)}
\includegraphics[angle=-90,width=0.3\linewidth]{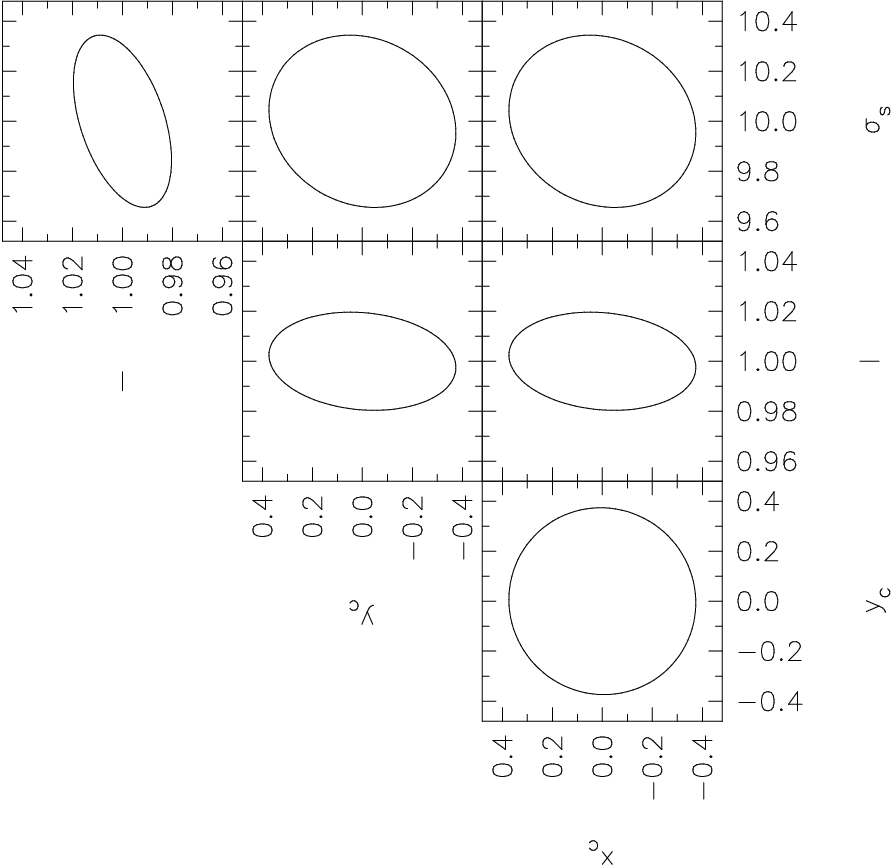}
\makebox[3mm][l]{\hspace{7mm}(f)}
\includegraphics[angle=-90,width=0.3\linewidth]{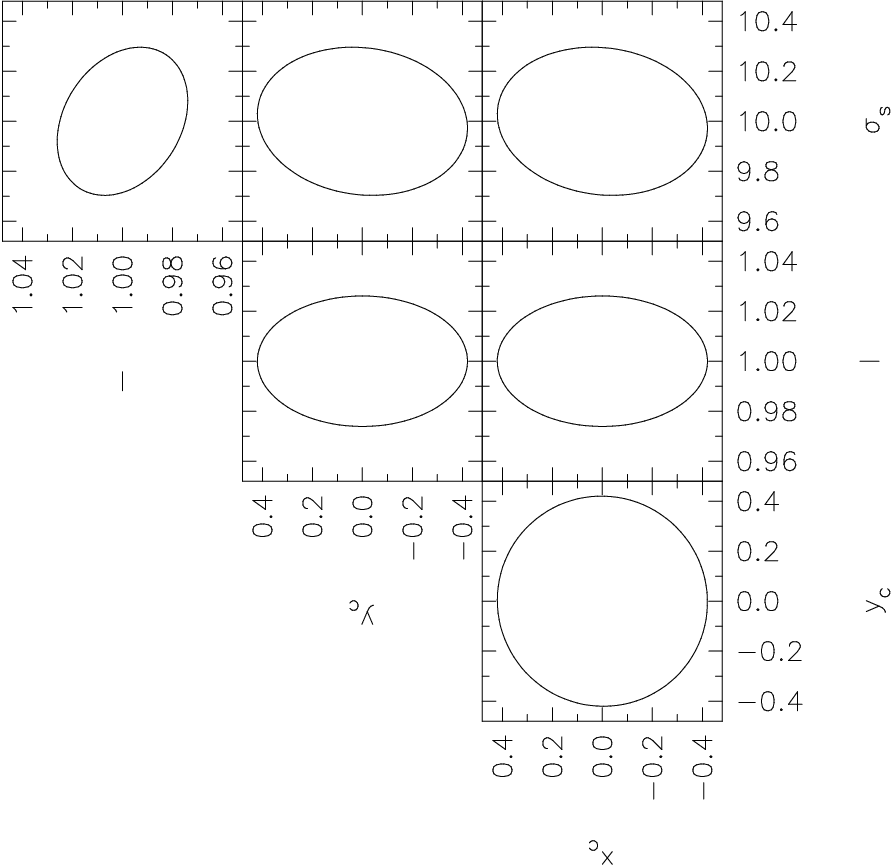}
\caption{\label{fig7}
Estimated 80\% marginalized confidence contours
for the recovered parameters of a Gaussian--Schell source, observed with
$17\times17$ close packed arrays of detectors of size 3 units and coherence 
10 units, but with
only $8 \times 8$ channels, so each channel is attached to 4 detectors.
In the top row (a--c), $2\times 2$ clusters of adjacent detectors are 
joined to the same channel, while in the bottom row (d--f) 
the detectors on each
channel are interlaced, such that there is one interposing detector
between detectors on the same channel. 
In all cases
the source has parameters $x_c=0$, $I=1$, $\sigma_s=20$.
The source has coherence length $\sigma_g=0$ (a, d), $\sigma_g=10$ (b, e)
and $\sigma_g=30$ (c, f).
}
\end{center}
\end{figure}

\subsection{Reduced channel arrays: effect of interlacing detectors
on the different channels}
In a real imaging array, there is a multiplexing problem, and so
it is often desirable to make the number of
channels smaller than the number of detectors, by connecting multiple
detectors to the same channel, such that the output of the channel
is the sum of the outputs of the individual detectors. 
The question then
arises: given a certain number of detectors and a certain number of channels,
what is the best way to group the detectors feeding into each channel?
The most obvious configuration is to have contiguous blocks of
detectors on the same channel, but the complexity of behaviour
we have seen in our previous simulations suggests that in some circumstances,
it may be desirable to interlace the detectors on the different channels.

We consider a close packed $17\times 17$ array of detectors of size 3 units
and coherence length 10 units, but with 4 detectors attached to each channel.
In the top row of Figure \ref{fig7} we show the recovered parameters
of sources of different coherence lengths where the 4 detectors
on each channel form a $2\times 2$ contiguous block. In the
bottom row, the detectors on the different channels are interlaced
such that the 4 detectors on each channel are at the corners of a $3 \times 3$
block.  

For the incoherent source, all the parameters are less well
constrained for the interlaced channels than the contiguous ones;
for the partially coherent source ($\sigma_g=10$), the parameters
are better constrained for the interlaced channels, and for the fully
coherent source ($\sigma_g=30$), the position and width are worse
constrained, but the intensity is better constrained, for the
interlaced array. The effects are quite small because the arrays
remain close packed, and the ``effective pixel size'' varies only from
6 to 9 units; Figures \ref{fig2} to \ref{fig5} would suggest
a relatively small effect over this range. However, it is possible
that with many detectors on each channel, and with the detectors on each
channel spread over a wider area, these effects may be amplified.
The complexity introduced makes further exploration beyond the scope
of this paper; however it is a potentially important area for future work.

\section{Conclusion}
We have combined our coupled--mode theory of detection with Fisher
information techniques from signal processing to study the performance
of imaging arrays. Several of our results have important implications
for array design. 
We have found that there is a threshold packing density above which
the detectors must be packed for information to be recovered, 
and that this threshold has little
dependence on the size and coherence of the source. Above and below
this threshold, the packing density makes relatively little difference.
As long as the source is not much smaller than the
detector spacing, the same information can be obtained regardless
of its position, even for sparse arrays.
On the other hand, the coherence length of both source and
detectors makes a significant difference to the amount of information that can
be obtained. Most importantly of all, we have shown that detailed
simulations are necessary for understanding the behaviour
of imaging arrays. The reason for this is that the behaviour results from 
the coupling between
the natural modes of the array and the natural modes of the source; when the
source and detectors are multimode, the behaviour thus becomes too complicated
to be explained by straightforward physical arguments.

Numerous extensions to our work are possible. One could study
the effects of photon occupancy for instance, or use a more sophisticated
model for the detectors. One could consider more complicated array geometries,
and explore the issue of interlaced pixels in more detail: our
results in this paper are inconclusive, but enough to suggest that interlaced
pixels may be preferable in certain conditions. Perhaps most interestingly,
one could investigate whether putting the outputs of
the detectors through a correlator enables more information
to be recovered, and can be used to mitigate the poor signal to noise
ratios for detectors placed in highly coherent fields.

\appendix

\section{Alternative definition of the Fisher matrix\label{APPA}}
The definition of the Fisher matrix in (\ref{fisherdef}) is that used by
most astronomers; however there is an equivalent definition 
\citep{kendall1967:1}, that is necessary to derive the Cram\'er--Rao bound.
Define the score vector $\bs{V}[\bs{\theta},\mathcal{H}]$ by
\begin{eqnarray}\label{score}
V_i[\bs{\theta},\mathcal{H}] &\equiv& \frac{\partial}{\partial\theta_i}
\ln\prob(\bs{D}|\bs{\theta},\mathcal{H})  \;.
\end{eqnarray}
We now show that the Fisher matrix is equivalent to the covariance of
the score.
From the definition of the score in (\ref{score}) we can show trivially that
\begin{eqnarray}
\expect[V_i] &=& 0 \;, \label{zeromean}
\end{eqnarray}
and therefore the covariance is given by
\begin{eqnarray}
\textrm{Cov} [V_i[\bs{\theta},\mathcal{H}], V_j[\bs{\theta},\mathcal{H}]]
&=& \expect\left[
\frac{\partial\ln\prob(\bs{D}|\bs{\theta},\mathcal{H})}{\partial\theta_i}
\frac{\partial\ln\prob(\bs{D}|\bs{\theta},\mathcal{H})}{\partial\theta_j}
\right] \nonumber \\
&=& \int \prob(\bs{D}|\bs{\theta},\mathcal{H}) 
\frac{\partial\ln\prob(\bs{D}|\bs{\theta},\mathcal{H})}{\partial\theta_i}
\frac{\partial\ln\prob(\bs{D}|\bs{\theta},\mathcal{H})}{\partial\theta_j}
\dif \bs{D} \nonumber \\
&=& \int \frac{1}{\prob(\bs{D}|\bs{\theta},\mathcal{H})}
\frac{\partial\prob(\bs{D}|\bs{\theta},\mathcal{H})}{\partial\theta_i}
\frac{\partial\prob(\bs{D}|\bs{\theta},\mathcal{H})}{\partial\theta_j}
\dif \bs{D} \;. \label{covexpand}
\end{eqnarray}
The Fisher matrix, as defined in (\ref{fisherdef}), is given by
\begin{eqnarray}
F_{ij}[\bs{\theta},\mathcal{H}] &=& 
- \int \prob(\bs{D}|\bs{\theta},\mathcal{H})
\frac{\partial^2\ln\prob(\bs{D}|\bs{\theta},\mathcal{H})}
{\partial\theta_i\partial\theta_j}
\dif \bs{D} \nonumber \\
&=& - \int \frac{\partial^2\prob(\bs{D}|\bs{\theta},\mathcal{H})}
{\partial\theta_i\partial\theta_j}
\dif \bs{D} 
+ \int \frac{1}{\prob(\bs{D}|\bs{\theta},\mathcal{H})} 
\frac{\partial\prob(\bs{D}|\bs{\theta},\mathcal{H})}{\partial\theta_i}
\frac{\partial\prob(\bs{D}|\bs{\theta},\mathcal{H})}{\partial\theta_j}
\dif \bs{D} \;. \nonumber \\ &&  \label{fishexpand}
\end{eqnarray}
But 
\begin{eqnarray}
\int \frac{\partial^2\prob(\bs{D}|\bs{\theta},\mathcal{H})}
{\partial\theta_i\partial\theta_j}
\dif \bs{D} = \frac{\partial^2}{\partial\theta_i\partial\theta_j}
\int \prob(\bs{D}|\bs{\theta},\mathcal{H}) \dif \bs{D} 
&=& 0 \;, 
\end{eqnarray}
so by comparing (\ref{fishexpand}) with (\ref{covexpand}) we find that
\begin{eqnarray}\label{fisheralt}
F_{ij}[\bs{\theta},\mathcal{H}] &=& 
\textrm{Cov} [V_i[\bs{\theta},\mathcal{H}], V_i[\bs{\theta},\mathcal{H}]] \;,
\end{eqnarray}
which is the alternative definition of the Fisher matrix.

\section{Derivation of the multi--parameter Cram\'er--Rao bound\label{APPB}}
Consider a statistic vector $\bs{T} = \bs{t}(\bs{D})$,
 that is an unbiased estimator of the parameter vector $\bs{\theta}$, i.e.
\begin{eqnarray}
\expect[\bs{T}] &=& \bs{\theta} \;.
\end{eqnarray}
Let $\bs{C}$ be the covariance matrix of the estimators:
\begin{eqnarray}
\bs{C} &\equiv& \expect[(\bs{T}-\bs{\theta})(\bs{T}-\bs{\theta})^\trans]
\end{eqnarray}
Now, using the definition of the score in (\ref{score}), we have
\begin{eqnarray}
\expect[V_iT_j] &=& \int T_j \frac{\partial}{\partial\theta_i} \prob(\bs{D}|\bs{\theta}) \dif \bs{D} \nonumber \\
&=& \frac{\partial}{\partial\theta_i}\expect[T_j] \nonumber \\
&=& \delta_{ij} \;. \label{evt}
\end{eqnarray}
Now form the matrix $\expect[(\bs{T}-\bs{\theta}-\bs{WV})
(\bs{T}-\bs{\theta}-\bs{WV})^\trans$, where $\bs{W}$ is some arbitrary 
matrix. By forming a quadratic form with some arbitrary vector $\bs{x}$, 
it is clear that
\begin{eqnarray} \label{start}
\expect[(\bs{T}-\bs{\theta}-\bs{WV})
(\bs{T}-\bs{\theta}-\bs{WV})^\trans] &\geq& \bs{0} \;,
\end{eqnarray}
i.e. the matrix on the left of the inequality is positive semi--definite.
Expanding the left hand side of (\ref{start}), 
using (\ref{fisheralt}), (\ref{zeromean})
and (\ref{evt}), and rearranging, we obtain
\begin{eqnarray} \label{fam}
\bs{C} &\geq&  \bs{W} +\bs{W}^\trans - \bs{WFW}^\trans \;.
\end{eqnarray}
Each matrix $\bs{W}$ gives us a different lower bound: to find the
maximum lower bound we diagonalize the Fisher matrix
$\bs{F}=\bs{Q}\bs{\Lambda}\bs{Q}^\trans$. It can then be easily shown that
\begin{eqnarray}
\bs{W} +\bs{W}^\trans - \bs{WFW}^\trans &=&
\bs{Q\Lambda Q}^\trans - (\bs{WQ}-\bs{Q\Lambda}^{-1})\bs{\Lambda}
(\bs{WQ}-\bs{Q\Lambda}^{-1})^\trans \;.
\end{eqnarray}
Hence the maximum lower bound is obtained for $\bs{W} = \bs{F}^{-1}$.
Substituting back in (\ref{fam}) we obtain
\begin{eqnarray}
\bs{C} &\geq& \bs{F}^{-1} \;.
\end{eqnarray}

\section{Fisher matrix for Gaussian likelihood\label{APPC}}
The Gaussian approximation to the likelihood function is
\begin{eqnarray}
\prob(\bs{D}|\bs{\theta},\mathcal{H}) &=&
\frac{1}{(2\pi)^{M/2}|\bs{\Sigma}[\bs{\theta},\mathcal{H}]|^{1/2}} 
\exp\left(-\frac{1}{2}(\bs{D}-\bs{\mu}[\bs{\theta},\mathcal{H}])^\trans
\bs{\Sigma}^{-1}[\bs{\theta},\mathcal{H}]
(\bs{D}-\bs{\mu}[\bs{\theta},\mathcal{H}])
\right) \;,
\end{eqnarray}
where $M$ is the number of parameters in the model.
Henceforth, for brevity, 
we do not show the dependence of $\bs{\mu}$ and $\bs{\Sigma}$ on the model
and its parameters explicitly in 
our notation.
We can thus write
\begin{eqnarray}\label{log}
\ln\prob(\bs{D}|\bs{\theta},\mathcal{H}) &=&
- \frac{1}{2}(\bs{D}-\bs{\mu})^\trans\bs{\Sigma}^{-1}(\bs{D}-\bs{\mu})
-N/2\ln(2\pi) - \frac{1}{2}\ln|\bs{\Sigma}| \;, \\
\label{1stdif}
\frac{\partial\ln\prob(\bs{D}|\bs{\theta},\mathcal{H})}{\partial\theta_i} &=&
\frac{\partial\bs{\mu}^\trans}{\partial\theta_i}
\bs{\Sigma}^{-1}(\bs{D}-\bs{\mu}) 
-\frac{1}{2}\frac{\partial\ln|\bs{\Sigma}|}{\partial\theta_i} 
- \frac{1}{2}(\bs{D}-\bs{\mu})^\trans
\frac{\partial\bs{\Sigma}^{-1}}{\partial\theta_j}(\bs{D}-\bs{\mu})
\end{eqnarray}
From (\ref{fisherdef}) the Fisher matrix is given by
\begin{eqnarray}
F_{ij}[\bs{\theta},\mathcal{H}] &=& 
-\expect\left[\frac{\partial^2\ln\prob(\bs{D}|\bs{\theta},\mathcal{H})}
{\partial\theta_i\partial\theta_j} 
\right]
\nonumber \\
&=&
\frac{\partial\bs{\mu}^\trans}{\partial\theta_i}\bs{\Sigma}^{-1}
\frac{\partial\bs{\mu}}{\partial\theta_j} 
+\frac{1}{2}\frac{\partial^2\ln|\bs{\Sigma}|}{\partial\theta_i\partial\theta_j}
+\frac{1}{2}\trace\left(\bs{\Sigma}\frac{\partial^2\bs{\Sigma}^{-1}}{\partial\theta_i\partial\theta_j}\right)
\end{eqnarray}

For the parameter values that maximize the likelihood, 
we set the expectation of the
first derivative equal to zero, to obtain
\begin{eqnarray} \label{logdet}
\frac{\partial\ln|\bs{\Sigma}|}{\partial\theta_i} &=& 
-\trace\left(\bs{\Sigma}\frac{\partial\bs{\Sigma}^{-1}}{\partial\theta_i}
\right)
\end{eqnarray}
and thus
\begin{eqnarray}
F_{ij}[\bs{\theta},\mathcal{H}]&=&\frac{\partial\bs{\mu}^\trans}{\partial\theta_i}\bs{\Sigma}^{-1}
\frac{\partial\bs{\mu}}{\partial\theta_j}
- \trace\left(\frac{\partial\bs{\Sigma}}{\partial\theta_j}
\frac{\partial\bs{\Sigma}^{-1}}{\partial\theta_i}\right)\;.
\end{eqnarray}
Finally we can write
\begin{eqnarray}
\trace\left(\frac{\partial\bs{\Sigma}}{\partial\theta_j}
\frac{\partial\bs{\Sigma}^{-1}}{\partial\theta_i}\right) &=&
\trace\left(\frac{\partial\bs{\Sigma}}{\partial\theta_j}
\frac{\partial(\bs{\Sigma}^{-1}\bs{\Sigma\Sigma}^{-1})}
{\partial\theta_i}\right)
\nonumber \\ &=& 
2\trace\left((\frac{\partial\bs{\Sigma}}{\partial\theta_j}
\frac{\partial\bs{\Sigma}^{-1}}{\partial\theta_i}\right) 
+ \trace\left(
\bs{\Sigma}^{-1} \frac{\partial\bs{\Sigma}}{\partial\theta_i}
\bs{\Sigma}^{-1} \frac{\partial\bs{\Sigma}}{\partial\theta_j}
\right)
\nonumber \\ &=&
- \trace\left(
\bs{\Sigma}^{-1} \frac{\partial\bs{\Sigma}}{\partial\theta_i}
\bs{\Sigma}^{-1} \frac{\partial\bs{\Sigma}}{\partial\theta_j}
\right) \;,
\end{eqnarray}
and thus we obtain (\ref{fisherapprox}).

\bibliography{/home/gs262/PhD/mybib}

\end{document}